\documentclass[sigconf,authorversion,nonacm]{acmart}

\usepackage[normalem]{ulem}
\usepackage{color}
\usepackage{float}
\usepackage{subcaption}
\usepackage{xspace} 
\usepackage{multirow}
\usepackage{array}
\usepackage{enumitem}
\usepackage[most]{tcolorbox}

\definecolor{burntorange}{rgb}{0.8, 0.33, 0.0}
\definecolor{byzantium}{rgb}{0.44, 0.16, 0.39}
\definecolor{byzantine}{rgb}{0.74, 0.2, 0.64}
\definecolor{lightlightgray}{rgb}{0.94, 0.94, 0.94}
\definecolor{lightblue}{rgb}{0.91, 0.95, 0.99}

\tcbset{on line, 
        boxsep=0pt,
        left=4pt,
        right=4pt,
        top=3pt,
        bottom=3pt,
        colframe=white,
        colback=lightblue,  
        highlight math style={enhanced}
        }

\AtBeginDocument{%
  \providecommand\BibTeX{{%
    \normalfont B\kern-0.5em{\scshape i\kern-0.25em b}\kern-0.8em\TeX}}}

\copyrightyear{2024}
\acmYear{2024}
\setcopyright{rightsretained} 
\acmConference[CHI '24]{Proceedings of the CHI Conference on Human Factors in Computing Systems}{May 11--16, 2024}{Honolulu, HI, USA}
\acmBooktitle{Proceedings of the CHI Conference on Human Factors in Computing Systems (CHI '24), May 11--16, 2024, Honolulu, HI, USA}
\acmDOI{10.1145/3613904.3642149}
\acmISBN{979-8-4007-0330-0/24/05}

\newcommand{\revision}[1]{{\color{black}#1}}
  \newcommand{\removed}[1]{}

\newcommand{\bolderandunderline}[1]{\textbf{\underline{#1}}}

\newcommand{\systemname}{Selenite\xspace} 
\newcommand{\exampleScenarioUser}{Mirri\xspace} 
\newcommand{\systemnameInFull}{\bolderandunderline{S}mart 
\bolderandunderline{E}nvironment for 
\bolderandunderline{L}ogical 
\bolderandunderline{E}xtraction and
\bolderandunderline{N}avigation of 
\bolderandunderline{I}nformation using 
\bolderandunderline{T}echnological
\bolderandunderline{E}xpertise\xspace}

\newcommand{\unakite}{Unakite\xspace}
\newcommand{\strata}{Strata\xspace}
\newcommand{\crystalline}{Crystalline\xspace}

\newcommand{\javascript}{JavaScript\xspace}

\newcommand{\userquote}[1]{``\textit{#1}''}

\newcommand{\criterionName}[1]{\tcbox{{\texttt{#1}}}}
\newcommand{\criterionNameSmall}[1]{\tcbox[left=2pt,right=2pt,top=1pt,bottom=1pt]{\small{\texttt{#1}}}}

\newcommand{\gptFour}{\texttt{GPT-4}\xspace}
\newcommand{\gptTwo}{\texttt{GPT-2}\xspace}

\begin{document}

\title{\systemname: Scaffolding Online Sensemaking with Comprehensive Overviews Elicited from Large Language Models}

\author{Michael Xieyang Liu}
\authornote{The author is currently affiliated with Google Research and can be contacted at lxieyang@google.com.}
\affiliation{%
  \institution{Carnegie Mellon University}
  \country{}
  }
\email{xieyangl@cs.cmu.edu}

\author{Tongshuang Wu}
\affiliation{%
  \institution{Carnegie Mellon University}
  \country{}
  }
\email{sherryw@cs.cmu.edu}

\author{Tianying Chen}
\affiliation{%
  \institution{Carnegie Mellon University}
  \country{}
  }
\email{tianyinc@andrew.cmu.edu}

\author{Franklin Mingzhe Li}
\affiliation{%
  \institution{Carnegie Mellon University}
  \country{}
  }
\email{mingzhe2@cs.cmu.edu}

\author{Aniket Kittur}
\affiliation{%
  \institution{Carnegie Mellon University}
  \country{}
  }
\email{nkittur@cs.cmu.edu}

\author{Brad A. Myers}
\affiliation{%
  \institution{Carnegie Mellon University}
  \country{}
  }
\email{bam@cs.cmu.edu}


\renewcommand{\shortauthors}{Liu and Wu et al.}

\begin{abstract}

Sensemaking in unfamiliar domains can be challenging, demanding considerable user effort to compare different options with respect to various criteria. Prior research and our formative study found that people would benefit from reading an overview of an information space upfront, including the criteria others previously found useful.
However, existing sensemaking tools struggle with the ``cold-start'' problem --- it not only requires significant input from previous users to generate and share these overviews, but such overviews may also turn out to be biased and incomplete. 
In this work, we introduce a novel system, \systemname, which leverages Large Language Models (LLMs) as reasoning machines and knowledge retrievers to automatically produce a comprehensive overview of options and criteria to jumpstart users' sensemaking processes. 
Subsequently, \systemname also adapts as people use it, helping users find, read, and navigate unfamiliar information in a systematic yet personalized manner. 
Through three studies, we found that \systemname produced accurate and high-quality overviews reliably, significantly accelerated users' information processing, and effectively improved their overall comprehension and sensemaking experience.

\end{abstract}

\ccsdesc[500]{Human-centered computing~Interactive systems and tools}

\keywords{Human-AI Collaboration, Sensemaking, Large Language Models, Natural Language Processing}


\maketitle

%
%
\section{Introduction}

Whether it is parents delving into the vast sea of baby stroller choices or developers comparing different \javascript frontend frameworks, people frequently find themselves having to make sense of unfamiliar domains and performing comparisons to make informed decisions.
In these situations, people often have to iteratively find, read, collect, and organize large amounts of information about different options with respect to various criteria \cite{kittur_costs_2013,kittur_standing_2014,liu_unakite_2018}, which can become a messy and overwhelming experience \cite{chang_mesh_2020,liu_crystalline_2022}. 

One challenge lies within the initial reading process --- due to unfamiliarity with the topic, people may struggle to fully understand certain content or fail to recognize important aspects that should otherwise warrant their attention \cite{klingberg_overflowing_2009,plumlee_effect_2003} as they read, leading to a limited viewpoint or ultimately misguided decisions \cite{yaniv_when_2012,hall_engaging_2007}. 
For example, novice developers might not be aware of criteria crucial to the utility of a software library, such as its stability, community size and support, or ease of integration with existing codebases, resulting in making a sub-optimal choice.

\begin{figure*}[t]
\vspace{-2mm}
\centering
\includegraphics[width=0.95\textwidth]{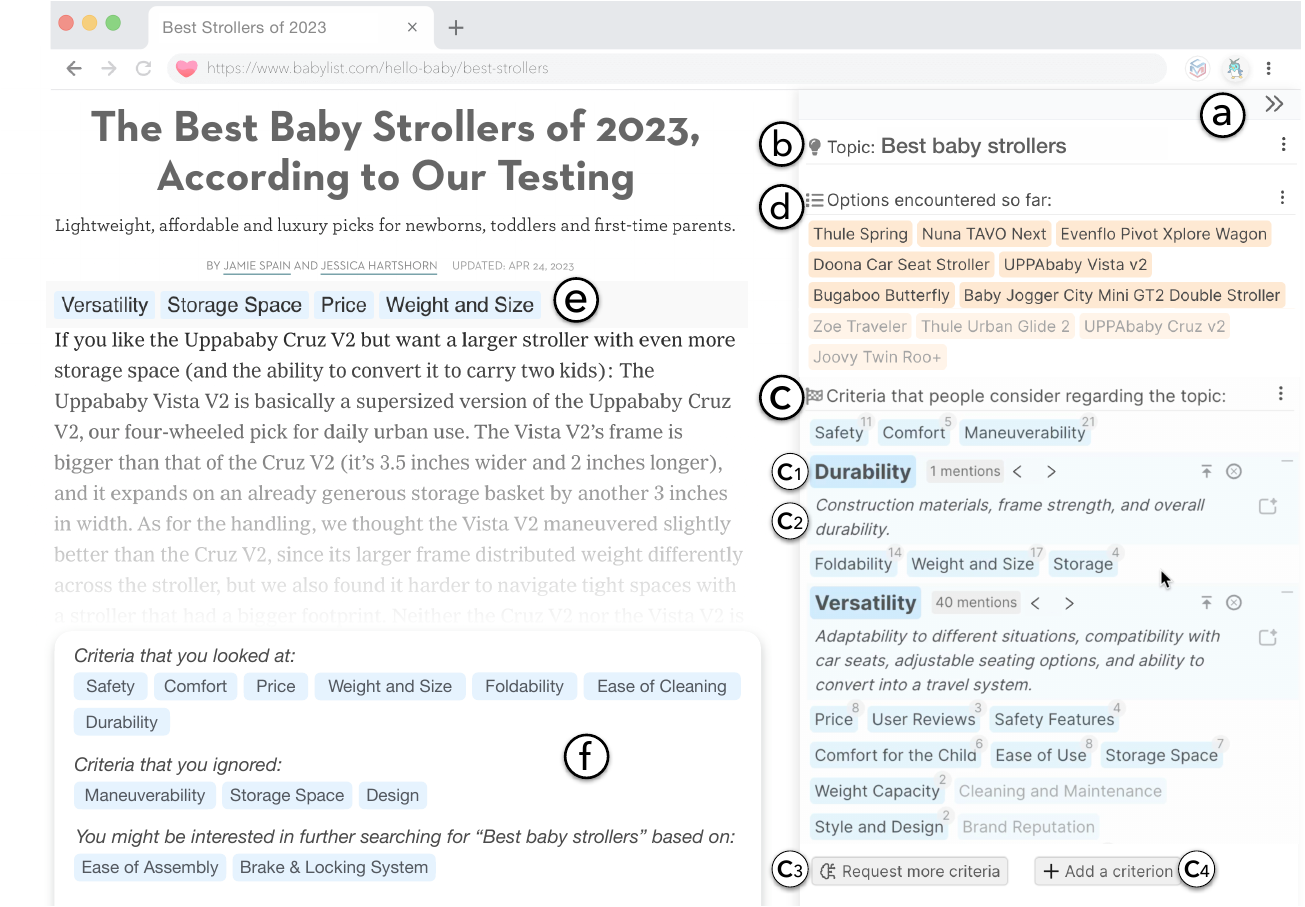}
\vspace{-3mm}
\caption{The main user interface of \systemname, which provides users with a comprehensive overview of the information space in the sidebar (a). When users encounter an unfamiliar topic (b), \systemname offers them a global overview based on commonly considered criteria (c) as well as the options encountered so far (d), helping them develop quick intuitions of the topic. As users read articles that they haven't seen before, \systemname provides local grounding through page-level and paragraph-level summaries and annotations (e), enabling effective comprehension and efficient navigation between the content of their interests. Before leaving a page, \systemname dynamically summarizes users' progress and suggests avenues for finding additional new information (f) in subsequent searches.
}
\vspace{-2mm}
\label{fig:selenite-sidebar}
\end{figure*}

Another challenge arises when people have to sift through numerous online reviews and comparison articles but with limited time and cognitive bandwidth, making a complete understanding of the information space impractical or impossible. 
Instead, people often adopt a ``selective'' (or non-linear) reading strategy \cite{willows_selective_1973}, only reading the paragraphs that discuss information that they consider relevant or valuable and bypassing the rest \cite{willows_reading_1974,cunningham_selective_1975}, e.g., in a focused session where they would like to compare different options with respect to the same criterion. However, it is challenging for people to gauge the potential value of articles or paragraphs, especially long-winded ones, just by skimming and without performing a more thorough read \cite{buder_selective_2015,farkas_designing_2013,fok_scim_2023}. 
For instance, information snippets about the maneuverability of different baby strollers may be dispersed throughout a review and appear in diverse variations (e.g., ``agile enough to go through tight spots'' and ``easy to steer and navigate small corners'' both discuss ``maneuverability''), making it difficult for people to spot these variations effectively and navigate efficiently among such scattered details. 

In response to these difficulties, prior work on sensemaking and knowledge reuse has shown promising evidence that people would benefit from seeing an \emph
{overview} of the information space \emph{before} they dive into the sensemaking process\cite{paul_cosense:_2009,morris_searchtogether:_2007,chang_searchlens_2019,suh_lifting_2008,fisher_distributed_2012} --- for example, Kittur et al. reported that having read an overview of the criteria that earlier users found useful can help people build intuition and understanding of the decision space upfront, leading to significantly improved digestion of the source material, better-structured mental models, and ultimately more well-informed decisions \cite{kittur_costs_2013,kittur_standing_2014}. Similarly, our formative study found that people expressed a desire for such comprehensive overviews to help them more systematically read, understand, and strategically navigate unfamiliar information. However, existing sensemaking systems have struggled with the ``cold start'' issue --- they often require substantial effort from the current user to personally go through the unfamiliar content and gather and structure information to obtain such an overview \cite{liu_unakite_2019,kang_threddy_2022,baldonado_sensemaker:_1997,schraefel_hunter_2002,kittur_costs_2013}, which defeats the premise of receiving one upfront. And even if a previous user has generated such an overview, it can often be incomplete \cite{paul_cosense:_2009,liu_reuse_2021}, biased \cite{flanagin_perceptions_2000,hoorn_web_2010}, or in idiosyncratic formats that make it hard for the current user to readily understand and leverage \cite{liu_unakite_2019,hsieh_exploratory_2018,horvath_understanding_2022,liu_tool_2023}.\looseness=-1


To overcome these challenges and go beyond prior systems, we explore the idea of providing users with a comprehensive overview of the information space upfront to jumpstart as well as guiding their subsequent sensemaking processes in a novel system named \systemname.\footnote{\systemname is named after a soft and transparent gemstone, and stands for ``\systemnameInFull.''} At a high-level, when users encounter an unfamiliar topic, \systemname leverages \gptFour, an LLM developed by OpenAI, as a knowledge retriever to offer them a global grounding based on commonly considered criteria, helping users develop quick intuitions of the topic (Figure \ref{fig:selenite-sidebar}c). As users read new articles, \systemname contextualizes that overview and uses it as an index to help users effectively comprehend and efficiently navigate among the content of their interest (Figure \ref{fig:selenite-sidebar}e).
Upon leaving a page, \systemname dynamically summarizes users' progress and suggests unique search queries that would help users find additional information and expand their perspectives rather than duplicating existing knowledge (Figure \ref{fig:selenite-sidebar}f). 
Through an intrinsic evaluation of \systemname, we verified its feasibility to provide a sufficiently accurate and high-quality global overview to users. Furthermore, additional usability and case studies revealed that \systemname accelerated users' information processing, facilitated their comprehension, and improved their overall reading and sensemaking experience. The contributions described in this work include:\looseness=-1

\begin{itemize}[leftmargin=0.2in]
    \item a formative study showcasing people's barriers and needs when reading and understanding information during online sensemaking, 
    despite recent advances in information management tools,

    \item \systemname, a novel system providing users with an upfront comprehensive overview of the information space as well as interactively guiding their subsequent reading and sensemaking processes,

    \item as part of \systemname, a novel user interface to interact with LLM-generated content and a novel user experience to contextualize that content into people's existing sensemaking workflows, going beyond the widely-adopted conversational interfaces in generative AI research and applications \cite{capra_how_2023,wondershare_chat_2023,blocktechnology_askyourpdf_2023,anthropic_claude_2023,openai_chatgpt_2023,ma_hypocompass_2023}.

    \item an intrinsic evaluation of \systemname that demonstrates the feasibility of our approach, as well as usability and case studies that offer preliminary insights into its usefulness and effectiveness,

    \item a discussion of design implications for future LLM and AI-powered sensemaking systems and tools.
\end{itemize}

%
%
\section{Related Work}\label{sec:rw}

\subsection{Making Sense of Online Information}
Building on theories of sensemaking as defined as developing a mental model of an information space in service of a user's goals \cite{dervin_overview_1983,klein_making_2006,russell_cost_1993}, at a high level, a typical online sensemaking process involves first \emph{reading and understanding} information from various data sources (e.g., online comparison articles, blog posts, video reviews, etc.), and then \emph{collecting and organizing} such information to form a schema or representation (e.g., a comparison table, a decision tree, etc.) to interpret the space \cite{pirolli_sensemaking_2005,chang_mesh_2020}. 
Notably, there have been many research as well as commercial tools and systems developed to support this latter stage of sensemaking, i.e., collecting and organizing information. For example, tools like SearchPad \cite{bharat_searchpad_2000}, Hunter Gather \cite{schraefel_hunter_2002}, as well as commercial systems like the Evernote clipper \cite{evernote_best_nodate}, allow a user to capture entire pages or portions of web content, categorize them, and later assemble them into a coherent document or structure for their own sensemaking, decision making, or sharing and collaboration. Furthermore, to reduce the disruption to users' overall workflow, prior work has also explored using lightweight interactions \cite{liu_wigglite_2022,chang_supporting_2016,chen_facilitating_2023,horvath_understanding_2022} to streamline the process of collection and triaging information, or even automatically keeping track of content of interest on behalf of the user \cite{dontcheva_summarizing_2006,liu_crystalline_2022,hogue_thresher_2005,chang_tabs_2021}.\looseness=-1

However, an inherent assumption that these prior systems make is that the user has \revision{a well-developed mental model that will allow them to effectively grasp the content} that they are about to collect using the system. But, in many circumstances, \revision{building sufficient context about the information space can be both time and effort-intensive, requiring individuals to iteratively read through and understand unfamiliar content by themselves without expert guidance or systematic strategy, as has been discussed in the literature \cite{nguyen_believe_2018,spichtig_decline_2016} and confirmed in our formative study. Despite being feature-rich, aforementioned sensemaking tools generally fail to address the ``cold start'' issue during the initial reading and comprehension stage where users need help deciding ``what is important to read'' \cite{fok_scim_2023} and ``how to most effectively read it'' \cite{august_paper_2023,rachatasumrit_citeread_2022,head_math_2022}.}
Discovering new and unseen content after reading can also be surprisingly difficult, as prior work revealed that a generic search query could return many redundant and near-duplicate documents \cite{pereira_where_2006,di_lucca_approach_2002,poonkuzhali_correlation_2011} despite the extensive use of duplicate detection algorithms in modern search engines \cite{plegas_reducing_2013}. Therefore, in this work, we focus on supporting users' reading and understanding of unfamiliar content in the first place. 

\revision{
\subsection{Tool Support for Reading and Comprehension}
One crucial way to address the ``cold start'' issue is to provide users with a summary of the information landscape. Previous research has investigated mechanisms that prompt domain experts to iteratively and collectively summarize a single document or discussion thread \cite{zhang_wikum:_2017,zhang_making_2018,gilmer_summit_2023}. In addition, prior work has introduced tools and systems that facilitate finding, extracting, and summarizing relevant information from multiple documents and sources into cohesive themes and knowledge \cite{liu_unakite_2019,chang_mesh_2020,chang_searchlens_2019}. However, these methods are dependent on the significant investment of time and effort from previous users and are contingent upon their availability, and may still result in summaries that are limited in scope \cite{paul_cosense:_2009,dourish_awareness_1992,liu_reuse_2021,zhang_making_2018} or with a biased perspective \cite{flanagin_perceptions_2000,hoorn_web_2010}. 

More recently, researchers have been experimenting with approaches to extract key information and build underlying knowledge structures automatically with machine learning \cite{das_development_2021,brody_unsupervised_2010,jin_opinionminer_2009,angelidis_summarizing_2018,fok_scim_2023}, lexical and HTML patterns \cite{kong_extending_2014,dontcheva_summarizing_2006,liu_crystalline_2022}, or crowdsourcing \cite{chang_alloy:_2016,chilton_cascade_2013,hahn_knowledge_2016}. However, many studies have demonstrated that these techniques can often produce structures that are incoherent and difficult for users to comprehend and contextualize \cite{chang_alloy:_2016,hearst_clustering_2006,chuang_interpretation_2012}. 

To address these limitations, \systemname leverages the wealth of world knowledge embedded within LLMs to generate overviews of information spaces that are aimed at being comprehensive and high-quality, eliminating the need for manual effort. In addition, \systemname provides detailed explanations of the generated overview as well as contextualizes it within the original content of the source documents to further enhance user understanding and facilitate navigation.
}

\subsection{Eliciting Knowledge from LLMs}
Recent advances in LLMs like \gptFour \cite{openai_gpt-4_2023}, PaLM \cite{chowdhery_palm_2022}, and LLaMa \cite{touvron_llama_2023} showcase impressive capabilities in answering user questions. These models are trained on large volumes of data, and as a result, their parameters might contain a significant body of factual as well as synthesized knowledge across a wide range of domains \cite{cohen_crawling_2023,wang_language_2020}, In this work, we leverage \gptFour as a knowledge retriever to retrieve a list of commonly considered aspects given an arbitrary topic, which is used to directly help users understand that topic at the beginning and systematically read about and explore that topic afterwards. 
However, LLMs face well-known challenges like hallucination and falsehood \cite{thorp_chatgpt_2023,bang_multitask_2023,terry_ai_2023}, which could make their outputs uncertain and less trustworthy, often requiring manual inspection and verification before use \cite{liu_what_2023,terry_ai_2023,ferdowsi_coldeco_2023,gordon_co-audit_2023}. In this work, we address these issues with a two-prong approach: 1) reducing hallucination through techniques such as Self-Refine \cite{madaan_self-refine_2023}; 
2) grounding LLM generations with the content that users would actually read, enabling natural verification.\looseness=-1

\section{Formative Study \& Design Goals}

To better understand the obstacles people encounter in their reading and sensemaking in unfamiliar domains, we first conducted a formative study. 

\subsection{Formative Study}

\subsubsection{Methodology}
Participants were a convenience sample of eight information workers (five male, three female) recruited through social media listings and mailing lists. To capture a variety of processes, we recruited three doctoral students, two professional software developers, two researchers, and one administrative staff member. While we do not claim that this sample is representative of all information workers, the interviews were very informative and helped motivate the design of \systemname.\looseness=-1

We began by asking participants to recall experiences of conducting sensemaking tasks on topics that they were not familiar with.\footnote{We subsequently kept track of these topics and used them in our system evaluations.} We then explored how they manage those situations. We asked participants to provide context by reviewing their browser histories to cue their recollections while retrospectively describing those tasks. We solicited their workflows, strategies, frustrations, and needs. Finally, we had participants use the open-source
\unakite system \cite{liu_unakite_2019}\footnote{\unakite is a Chrome extension that helps users collect and organize information into comparison tables in a lightweight fashion while searching and browsing web articles. We used \unakite since it has been shown to be easy to learn and use, and can support a diversity of web page styles and structures \cite{liu_unakite_2019}.} to make sense of a topic that they were not familiar with (e.g., for people who have not yet had children to figure out the best baby strollers to purchase for their future child) and externalize their workflows, strategies, and mental models.\looseness=-1 

\subsubsection{Findings}

Participants reported a total of 28 distinct topics that they encountered and explored, such as 
\userquote{choosing a hybrid app framework,} 
\userquote{selecting the best time tracking tool,} 
and \userquote{picking an engagement ring.}\footnote{For a complete catalog of these topics, please refer to Table \ref{tab:selenite-formative-topics} in the Appendix.}
Below, we report major findings from the study:

\textbf{People often find themselves feeling lost or unsure of where to begin and desire a big-picture understanding of important criteria (or aspects) of an information space before diving deeper.} When approaching an unfamiliar topic, one common strategy that participants reported employing is to find some sort of \userquote{overview of different aspects} (P3) that would give them \userquote{an intuition of what to care about and some guidance on what to look out for regarding each option} (P5) in their subsequent exploration. For example, when investigating which time-tracking app to use, P5 was able to find a few articles that provided such overviews at the beginning, e.g., under the section \userquote{What makes the best time tracking software?}
However, participants complained that such lists of criteria are often \userquote{subjective, incomplete} (P1), can contain aspects that they \userquote{most likely don't care about} (P2), and worse yet, \userquote{do not represent how the rest of an article would be structured} (P6). In addition, for certain topics such as \userquote{best birthday gift ideas}, such overviews of criteria are hard to find upfront, in which case participants would have to employ a bottom-up approach by reading through a series of articles back-and-forth, which is often considered \userquote{time-consuming} (P1) and \userquote{hard to actually follow through} (P2). Without these \userquote{important criteria to keep in mind} (P4) upfront, participants reported feeling \userquote{overwhelmed by large amounts of unfamiliar information} (P6), lacking \userquote{a sense of clarity and structure} (P7), and can easily lose focus during sensemaking. These findings prompted us to generate an initial overview of the commonly considered criteria given an information space to provide users with some global grounding and an anchor point for their subsequent reading and sensemaking.

\begin{figure*}[t]
\centering
\includegraphics[width=0.8\textwidth]{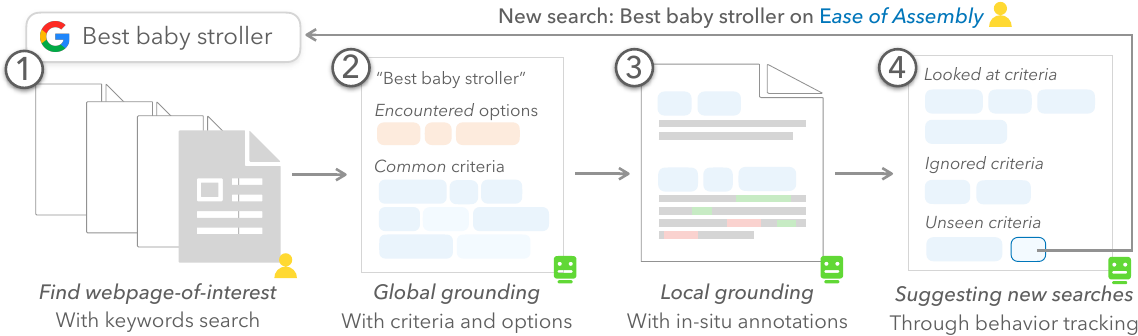}
\vspace{-2mm}
\caption{\revision{Main stages and features of \systemname: After the user 1) searches and finds an initial webpage-of-interest to read, \systemname provides: 2) global grounding with a set of common criteria as well as options encountered so far, 3) local grounding with in-situ annotations of criteria per paragraph, and 4) suggestions on what to search for next to gain new information.}}
\label{fig:selenite-system-architecture}
\end{figure*}

\textbf{Identifying and consistently keeping track of criteria is challenging. 
} \revision{Participants cited identifying and aggregating criteria while reading content as a \userquote{significant cognitive load} (P4).} \revision{One challenge is that the same criterion can be discussed in various ways across different articles (and even within the same article), making it hard for participants to recognize those variations and time-consuming to flip back and forth to make sure they are eventually aggregated and consistently represented in their \unakite tables.} For example, when investigating the topic of ``baby strollers,'' P6 first saw a stroller should be ``agile and nimble to be able to go through tight spots and sidewalks,'' and recorded it as ``nimbleness'' in \unakite; later when she saw another segment that stated that a particular stroller is ``easy to steer and handle and can smoothly navigate tight corners,'' she created another criterion called ``steering.'' It wasn't until when she saw a segment in a third article that described a stroller having great ``maneuverability and control'' did she realize that all of these were practically describing the same aspect, ``maneuverability,'' and she had to go back and readjust and combine those criteria and their associated evidence in \unakite, \revision{a common refactoring challenge for users in unfamiliar domains \cite{kittur_standing_2014}.} 
Additionally, P7 recounted a similar experience when searching for washers and dryers for his first house and admitted that \userquote{oftentimes, coming up with the right keyword or jargon to summarize what I saw can be surprisingly hard, and I really wish someone would just do that for me.} 
\systemname tackles this issue by providing a comprehensive list of frequently considered criteria upfront, \revision{reducing the need for individuals to haphazardly find and aggregate criteria themselves}.


\textbf{People need reading and navigation guidance both at paragraph level as well as article level.} Participants reported often having \userquote{limited attention span} (P8) when reading online articles and can only focus on a certain amount of information, usually the first few paragraphs or the first few sentences within a paragraph, before getting distracted or lost. For example, P5 in her quest to find a suitable time-tracking app pointed to a typical situation where \userquote{sometimes a paragraph, even a short one, could be quite convoluted and have a lot of intertwined information, for example, and at first I thought this paragraph was just about money, but the rest of the paragraph was actually about lots of other things like platform compatibility.} But, since participants tend to skim through content quickly, it often leads to potential misunderstandings or missing important details. In such situations, participants desired \userquote{some simple metadata of what's covered in a paragraph} (P4) to give them an intuition of what the paragraph is about and whether it is worth reading. These findings prompted us to provide in-situ per-paragraph summaries and the option for users to clarify convoluted paragraphs by ``zooming in'' on them in \systemname.

The same applies to the page level, where participants wanted to be able to \userquote{preview a page before investing time reading it} (P3) to understand whether it discusses detailed aspects that they care about. Additionally, such preview can also help them \userquote{maximize the information gained from each page} (P5), i.e., help them avoid reading duplicate information and aspects without learning anything new. As P4 put it, \userquote{if I've already learned about all the aspects from the other pages, I don't have to read this one.} However, as discussed previously, such previews (even if they are in the form of an abstract or table of contents) are not always available for each article, \revision{nor grounded in a person's past reading and information collection activity.} \systemname addresses the page-preview need by offering users a concise overview of what's covered (and not) in a page to help users gauge its value. Additionally, \systemname tackles the issue of personalization by presenting users with a progress summary based on their previous sensemaking activities at the end of a page.\looseness=-1

As participants became more familiar with a topic, their reading patterns started to get increasingly selective and non-linear. For example, we have observed that participants use a combination of keyword searches and flipping back and forth in an effort to find relevant information about a particular criterion that they cared about (with respect to different options), which they thought was \userquote{haphazard} (P1) and \userquote{inefficient} (P7). This led us to suggest potentially fruitful search keywords to users for discovering more unseen information in the end-of-page progress summary.\looseness=-1

\subsection{Summary of Design Goals}
We postulate that an effective user interface/interaction paradigm for helping users find and read about key information during sensemaking should support:

\begin{itemize}[leftmargin=.15in]
    \item \textbf{[D1]} As the user starts investigating a topic, \textbf{provide a \emph{global grounding} using common criteria as well as the options encountered} to help users build intuitions of the information space and promote structured thinking; 
    
    \item \textbf{[D2]} During their reading, \textbf{provide a \emph{local grounding} using page-level as well as paragraph-level summary and annotation} to enable an accurate understanding of and effective navigation within and across articles; 

    \item \textbf{[D3]} Upon finishing, \textbf{dynamically suggest next steps in sensemaking based on users' existing reading and information collection activities} to avoid missing important aspects \emph{after reading} as well as maximize the information gain \emph{in future readings}. 
\end{itemize}

\section{The \systemname System}

Based on the design goals, we designed and implemented the \systemname Chrome extension prototype to help people read about and make sense of unfamiliar topics with the help of global grounding \revision{(summarized in Figure \ref{fig:selenite-system-architecture})}. We will first illustrate how an end-user, \exampleScenarioUser, would interact with \systemname.

\subsection{Example Usage Scenario}
\exampleScenarioUser, an expectant mother, is seeking guidance in picking a stroller for her upcoming baby. As someone without prior experience in child-rearing, she decided to rely on \systemname to help her while going through review articles and product pages of baby strollers.\looseness=-1

\exampleScenarioUser did a quick Google search and clicked on the first result page, which appeared to be a review article titled ``The 10 Best Baby Strollers Put To The Test''. Upon opening the page, \systemname automatically recognized the \textbf{topic} of the page as ``best baby strollers'' (Figure \ref{fig:selenite-sidebar}b), \revision{and then automatically presented an overview in a global sidebar} \revision{available on every page} (Figure \ref{fig:selenite-sidebar}a). \revision{The overview contained a list of \textbf{criteria} (Figure \ref{fig:selenite-sidebar}c) that are commonly considered by people when investigating the topic, such as \criterionNameSmall{maneuverability} and \criterionNameSmall{durability}. In addition, \systemname also automatically parsed the web page content and extracted the different baby stroller \textbf{options} and presented them under the ``Options encountered so far'' section (Figure \ref{fig:selenite-sidebar}d) in the sidebar. 
After quickly skimming the overview, \exampleScenarioUser now felt that she has already built an intuition about what criteria she should care about when picking baby strollers before even delving into the article itself.}\looseness=-1

Based on the global options and criteria, \systemname \emph{contextualizes} the ones that are covered on the current page by highlighting them in the sidebar \revision{(and conversely low-lighting the ones that are not present, for example, see Figure \ref{fig:selenite-sidebar}c\&d),} helping users better understand and find specific information of interest while browsing. 
In addition, as \exampleScenarioUser read the article, \revision{she noticed that \systemname provided \textbf{\revision{in-situ} annotations} of \textbf{mentioned criteria} at the beginning of each paragraph 
(Figure \ref{fig:selenite-sidebar}e).} 
She quickly learned that she could just skim those mentioned criteria to get a rough idea of what a particular paragraph is about and decide if that paragraph is worth reading.\looseness=-1

\begin{figure}[t]
\vspace{-2mm}
\centering
\includegraphics[width=0.99\linewidth]{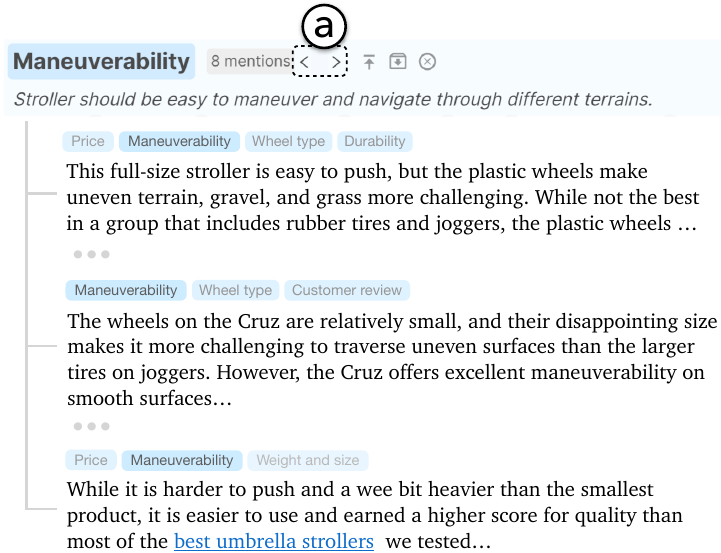}
\vspace{-2mm}
\caption{\revision{\systemname enables structured and efficient navigation by criterion through clicking the ``previous/next'' (shown as ``<'' and ``>'') buttons (a), after which \systemname will automatically scroll the page to reveal the previous/next mentioning of the target criterion.}}
\label{fig:selenite-navigate}
\vspace{-3mm}
\end{figure}

When she came across information about the \criterionNameSmall{maneuverability} of a specific stroller, 
\exampleScenarioUser became interested in finding out if there were any details about the maneuverability of other stroller options as well. To facilitate this, she used the ``previous/next'' buttons (Figure \ref{fig:selenite-navigate}a) to quickly navigate among the paragraphs that discussed maneuverability. Here, the aforementioned annotations not only offer paragraph overviews during \emph{linear} skimming but also act as bookmarks for \emph{non-linear} navigation between distinct parts of the page that pertain to similar criteria.\looseness=-1

Later, when \exampleScenarioUser encountered a particularly convoluted paragraph with multiple criteria and options that she couldn't quite absorb after a first pass, she decided to leverage the ``zoom in'' feature that \systemname offers --- querying for more comprehensive descriptions that clarify which sentences or phrases within the paragraph pertain to specific criteria and sentiments (positive, neutral, or negative) (Figure \ref{fig:selenite-zoom-in}) by clicking the ``Analyze'' button (Figure \ref{fig:selenite-zoom-in}a) that appears when hovering the cursor over a paragraph.

\revision{
After \exampleScenarioUser reached the end of the current article, \systemname presented a \textbf{summary} block (Figure \ref{fig:selenite-sidebar}f), automatically summarizing her research progress (e.g., criteria from the overview that she has actually read about). With the help of this summary, \exampleScenarioUser realized that she hasn't seen evidence related to \criterionNameSmall{ease of assembly} or \criterionNameSmall{brake \& locking system} before. She then specifically searches for them on Google, finding new articles that contain information about these previously unencountered criteria.
}


\subsection{Detailed Designs}

We now discuss how the various \systemname features are designed and implemented to support the design goals.

\subsubsection{\textbf{[D1] Providing Global Grounding using Common Criteria and Options Encountered}}\label{sec:selenite-d1-detail}


In \systemname, we explore the idea of having the system provide users with an initial overview of criteria that are typically significant and frequently considered by people when exploring a particular topic. 
\systemname also performs information extraction on each page to identify the options that a user has encountered during their sensemaking process. Naturally, users have the flexibility to reorder, pin, edit, add, or delete any options and criteria to tailor them precisely to their specific preferences. We discuss the relevant designs and the rationale behind those designs below:

\begin{figure}[t]
\vspace{-2mm}
\centering
\includegraphics[width=1.02\linewidth]{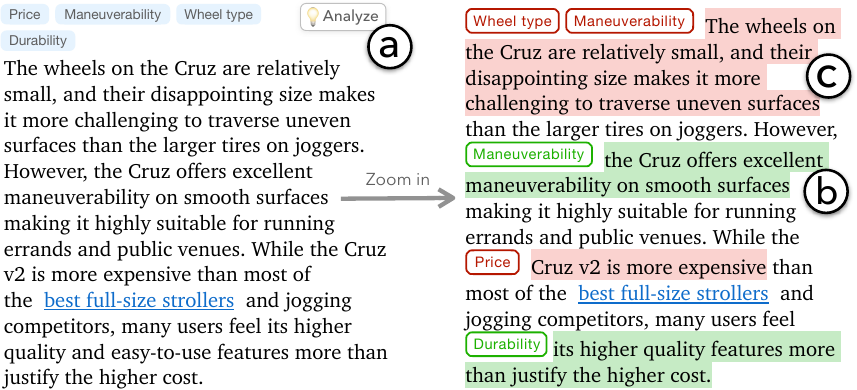}
\vspace{-7mm}
\caption{When encountering a particularly convoluted paragraph (e.g., the paragraph on the left) with multiple criteria and options that users can't quite absorb in the first pass, they can click the ``Analyze'' button (a) and leverage the ``zoom in'' feature that \systemname offers to query for more comprehensive descriptions that clarify which sentences or phrases pertain to which specific criteria and sentiments. \systemname wraps phrases and sentences in colored boxes, with green denoting ``positive'' (b), red denoting ``negative'', and grey denoting ``neutral'' (not shown).}
\vspace{-2mm}
\label{fig:selenite-zoom-in}
\end{figure}

\paragraph{Automatically recognizing topics.} \systemname goes beyond previous sensemaking systems \cite{liu_unakite_2018,chang_tabs_2021,kittur_standing_2014,liu_crystalline_2022,kittur_can_2008,liu_wigglite_2022,liu_reuse_2021} by autonomously identifying and classifying web pages into broad topics based on their titles and content. 
Unlike previous systems that require users to \revision{manually determine the topic \cite{chang_mesh_2020,liu_unakite_2019,hahn_bento_2018}, automatic topic recognition further lowers the barrier for entry, enabling users to quickly begin reaping benefits from the list of commonly considered criteria based on that topic.}
To achieve this, we frame the topic recognition as a \emph{summarization} task for an LLM ---
\revision{specifically, we asked \gptFour (taking advantage of its generalizability to various domains \cite{openai_gpt-4_2023})\footnote{Please refer to section \ref{sec:gpt-prompt-get-topic} in the Appendix for the detailed prompt design.} to first \emph{summarize} an arbitrary web page given its title and initial five paragraphs} (with the temperature set to 0 to minimize LLM hallucination).\footnote{Empirically, we found this step helped \gptFour to better engage with the context provided. It also aligns with the idea of Chain-of-thought prompting proposed by \cite{wei_chain--thought_2023}, and then offer a \emph{search phrase} that would help one to find similar web pages using a modern search engine, which we use as the topic.} \systemname then clusters the semantically similar topics (note that each web page has an associated topic generated by \gptFour) based on the cosine distances on topic semantic embeddings computed using SentenceBERT~\cite{reimers_sentence-bert_2019}. 
\revision{Therefore, for example, websites titled ``React vs. Svelte: Performance, DX, and more,''
``Angular vs React vs Vue: Which Framework to Choose,''
and ``What are the key differences between Meteor, Ember.js and Backbone.js?''
would all be regarded as ``Comparison of JavaScript frameworks.''}
Naturally, users have the flexibility to manually create, edit, and remove topics, as well as reassign pages to different topics based on their personal opinions.


\paragraph{Automatically retrieving commonly considered criteria.}
If we adopt the ``bottom-up'' approach discussed in prior work \cite{hahn_bento_2018,liu_unakite_2019,kuznetsov_fuse_2022}, an intuitive method for obtaining criteria would involve extracting them from individual paragraphs on a page. 
However, in our initial attempts, we found that this method faced significant challenges that limited its effectiveness. One of the main issues was the lack of uniformity among the criteria extracted from different paragraphs --- each paragraph presented its own variations and nuances, making it difficult to establish a cohesive and standardized set of criteria (similar to what was reported by our formative study participants). Additionally, the approach lacked a comprehensive global perspective, failing to consider the broader context and overarching themes of the topic. As a result, manual review, correction, and unification of the extraction results were frequently necessary, making the process impractical and inefficient.\looseness=-1

\def\arraystretch{1.2}
\begin{table*}[t]
\vspace{-2mm}
\centering
\resizebox{1\textwidth}{!}{%
\revision{
\begin{tabular}{
r 
p{168mm}
}
\toprule
\textbf{Criterion Name} & \textbf{Criterion Description} \\ 
\midrule

\criterionName{Safety} &
Ensuring the stroller has proper safety features such as a secure harness, sturdy construction, and reliable brakes. \\

\criterionName{Comfort} &
Providing a comfortable seat with adequate padding and support for the baby, as well as adjustable recline positions. \\

\criterionName{Maneuverability} &
Having easy \& smooth maneuverability, with features like swivel wheels, suspension systems, and the ability to navigate tight spaces. \\

\criterionName{Durability} &
Ensuring the stroller is built to last, with high-quality materials and strong construction. \\

\criterionName{Storage} &
Offering ample storage space for carrying essentials such as diaper bags, snacks, and personal items. \\

\criterionName{Folding and Portability} &
Allowing for easy folding and compact storage, as well as being lightweight for convenient transportation. \\

\criterionName{Versatility} &
Providing features that allow the stroller to adapt to different terrains, weather conditions, and age ranges. \\

\criterionName{Ease of Use} &
Having user-friendly features like adjustable handles, intuitive controls, and easy-to-clean fabrics. \\

\criterionName{Price} &
Considering the affordability and value for money in relation to the features and quality of the stroller. \\

\criterionName{Customer Reviews} &
Taking into account feedback and recommendations from other parents who have used the stroller. \\

\criterionName{Weight and size} &
Considering the weight and size of the stroller to ensure it is manageable and fits well in different environments. \\

\criterionName{Ease of cleaning} &
Ensuring the stroller is easy to clean and maintain, with removable and washable fabric components. \\

\criterionName{Adjustability} &
The stroller should have adjustable handlebars and footrests to accommodate different caregivers and growing babies. \\

\criterionName{Canopy} &
A large and adjustable canopy to protect the baby from the sun and other elements. \\

\criterionName{Reversible seat} &
Having the option to face the baby towards the parent or away from the parent. \\

\criterionName{Brake system} &
Having a reliable brake system that is easy to engage and disengage. \\

\criterionName{Car seats compatibility} &
Offering the ability to attach a car seat to the stroller for convenient travel. \\

\criterionName{Adjustable height} &
Allowing for adjustable handlebars to accommodate different heights of caregivers. \\

\criterionName{Easy assembly} &
Providing clear instructions and easy assembly process for the stroller. \\

\criterionName{Design and aesthetics} &
Considering the overall design and aesthetics of the stroller to match personal preferences. \\

\criterionName{Weight capacity} &
Specifying the maximum weight limit the stroller can safely carry. \\

\criterionName{Warranty} &
Checking for a warranty or guarantee that covers any potential defects or issues with the stroller. \\

\criterionName{Brand reputation} &
Considering the reputation and reliability of the brand manufacturing the stroller. \\

\criterionName{Accessories} &
Offering additional accessories such as rain covers, mosquito nets, or parent organizers for added convenience. \\
\bottomrule
\end{tabular}%
}
}
\vspace{0mm}
\caption{\revision{Commonly considered criteria (name \& description) that \systemname retrieves for the topic of ``best baby strollers.''}}
\label{tab:selenite-baby-stroller-sample-criteria}
\vspace{-5mm}
\end{table*}

To address these challenges, we instead explored an alternative ``top-down'' approach, where we directly query an ``oracle'' for a globally applicable and comprehensive set of criteria.
We were particularly inspired by recent research suggesting that \emph{the majority of people's information-seeking needs are not novel}~\cite{liu_evaluating_2023} --- ``previous people'' have experimented with most search needs and synthesized information into summarized knowledge such as review articles. 
While it is impractical for individuals to process and synthesize vast amounts of information online, LLMs excel at this. Recent studies suggest that LLMs can be highly effective in processing and integrating information, making them potentially valuable for tasks like knowledge graph querying and retrieving common sense information \cite{yang_empirical_2022,alkhamissi_review_2022,wang_language_2020}, and, in our particular case, a suitable ``oracle'' for providing a set of commonly considered criteria given a particular topic.\looseness=-1 

In \systemname, 
\revision{we use \gptFour as a \emph{knowledge retriever} --- for any given topic, we prompt it to produce a list of around 20 commonly considered criteria (Figure \ref{fig:selenite-sidebar}c), complete with their respective names (Figure \ref{fig:selenite-sidebar}c1) and descriptions (Figure \ref{fig:selenite-sidebar}c2) for each topic.}
\revision{To minimize potential anchoring biases, we strive to achieve a balance between relevance and diversity in our prompting strategy.} 
First, we specifically requested an initial set of criteria that are deemed as ``most relevant to the topic,'' ``frequently considered,'' and can ``cover a broad range of perspectives.'' 
\revision{Then, we iteratively prompted \gptFour while applying the Self-Refine technique~\cite{madaan_self-refine_2023}, where in each iteration, we requested the generation of five additional criteria that were ``different, more diverse, and more important'' than the previous ones.}\footnote{Please refer to section \ref{sec:gpt-prompt-get-criteria} in the Appendix for the detailed prompt design.} We also relied on \gptFour for ranking the criteria based on their importance.
Still, users have the freedom to request additional criteria (without repetition) if they believe the existing list is not comprehensive enough (Figure \ref{fig:selenite-sidebar}c3), or manually add criteria (Figure \ref{fig:selenite-sidebar}c4). 
\revision{We present in Table \ref{tab:selenite-baby-stroller-sample-criteria} the list of criteria that \systemname retrieves for the topic of ``best baby strollers,'' offering an intuition of their quality and coverage. 
We further validate our approach through a performance evaluation in section \ref{sec:selenite-performance-eval} that provides initial evidence that our approach is sufficient for our prototyping purposes. We leave for future work to experiment with advanced approaches, such as retrieval-augmented generation (RAG) \cite{lewis_retrieval-augmented_2020}, that would potentially provide increased perceived external validity.}




\paragraph{Automatically recognizing encountered options.} 
Instead of relying on \gptFour to access its internal knowledge and retrieve a set of commonly considered options, we instead leverage its zero-shot \emph{information extraction} capability
and expansive context window size \cite{openai_gpt-4_2023}
to directly extract options from the entire text content of a web page. This approach ensures that the options presented in the sidebar align with a user's sensemaking process, i.e., they are indeed what users have encountered as opposed to something that users would potentially never run into. It also circumvents the potential concern where the world knowledge of an LLM is out-of-date, for example, at the time of writing, \gptFour only ``knows'' information up to September 2021 \cite{openai_gpt-4_2023}.\footnote{While the direct retrieval of criteria from LLMs may also face this potential issue, in practice, we operate under the assumption that criteria are unlikely to suddenly emerge or become outdated.} In addition, it surpasses the limited heuristics employed in previous approaches such as \crystalline \cite{liu_crystalline_2022}, which rely on page titles and HTML \texttt{<h>}-tags as sources for options. 
This is crucial, because studies have consistently demonstrated that web pages frequently disregard semantic web standards and best practices \cite{mendes_toward_2018,henschen_using_2009}.\footnote{For instance, it is common to find pages where nearly every piece of content is enclosed in \texttt{<div>} tags regardless of their semantic roles.}\looseness=-1


\subsubsection{\textbf{[D2] Providing Local Grounding using Page \& Paragraph-level Summary and Annotation}}

\revision{
To address the challenges identified in the formative study, where complex paragraph and page structures frequently led to overlooked information and hindered user comprehension, \systemname introduces the following features:
}

\paragraph{\revision{In-situ} summaries and annotations of paragraphs.} With access to the initial set of common criteria as well as the options extracted from each page, \systemname performs content analysis on each paragraph within a given page to identify the specific criteria being discussed and presents them as \revision{in-situ} annotations above the respective paragraph (Figure \ref{fig:selenite-sidebar}e). This feature enables users to swiftly scan through a page, understand the key points of each paragraph, and selectively concentrate on the paragraphs that are valuable and engaging for gathering information.

Such content analysis is enabled by recent advances in large pre-trained transformer models \cite{vaswani_attention_2017,devlin_bert_2019,lewis_bart_2019} fine-tuned to perform zero-shot text classification tasks following a natural language inference (NLI) paradigm \cite{yin_benchmarking_2019}. Specifically, for example, to assess if a given text (e.g., ``\texttt{Angular is very hard to pick up}'') covers the criterion of \criterionNameSmall{learning curve}, we can input the text as the \emph{premise} and a \emph{hypothesis} of ``\texttt{This content discusses \{learning curve\}}.'' into the NLI model. The entailment and contradiction probabilities are then converted into label probabilities, indicating the likelihood that the content pertains to the specified criterion. We used the \texttt{bart-large-mnli} model\footnote{The model can be accessed on-demand through a remote API service that we implemented.} for this purpose and considered options and criteria with a score above 0.96 as true positives, displaying them in descending order of scores. We determined this threshold empirically, prioritizing recall over precision, as discussed further in section \ref{sec:selenite-performance-eval}.

In scenarios where users still struggle to comprehend content despite the presence of \revision{in-situ} annotations, \systemname can perform a deeper analysis on-demand by leveraging the advanced reasoning capabilities of \gptFour. Specifically, through parallel and carefully orchestrated prompts, \systemname produces a more comprehensive description that clarifies which sentences or phrases pertain to specific criteria and sentiments (positive, neutral, or negative) (Figure \ref{fig:selenite-zoom-in}). \revision{Although a formal evaluation of this method is beyond the scope of this work, recent research suggests that content analysis conducted by the latest generation of LLMs achieves state-of-the-art performance in terms of quality, accuracy, and granularity \cite{openai_gpt-4_2023,bubeck_sparks_2023}, making it suitable for our purposes.} 

\paragraph{Page-level overview of options and criteria.} 
As evidenced by our formative study, providing a page-level overview of the information space to users can greatly assist them in reading and sensemaking tasks. To facilitate this, \systemname consolidates paragraph-level metadata into the sidebar's options and criteria entries, with the entries that are present on the page highlighted (Figure \ref{fig:selenite-sidebar}c\&d). This offers two key benefits. First, it provides a comprehensive summary of all the available options and criteria specific to the current page, which allows users to quickly understand the focus of the page as well as judge its value against their personal interests and needs. Second, it enables structured and efficient navigation. By utilizing the ``previous/next'' button for a given criterion (Figure \ref{fig:selenite-navigate}a), users can swiftly navigate between distinct parts of the page related to identified criteria (Figure \ref{fig:selenite-navigate}). This feature saves users time and effort, as it eliminates the need for manual searching and filtering, which our formative study found to be the common practice.

It is worth noting that the combination of page and paragraph-level annotations \revision{effectively addresses a significant prior limitation highlighted by \crystalline \cite{liu_crystalline_2022} and further revealed in our formative study:} the inability to manually recognize ``latent/implicit criteria,'' where the same criterion can be expressed in various forms without being explicitly mentioned (for instance, it can identify the criterion of ``price'' from a statement like ``I bought this mp3 player for almost nothing'' \cite{poria_rule-based_2014}).\looseness=-1 

\subsubsection{[\textbf{D3] Dynamically Suggesting Next Steps in Sensemaking}}

\revision{
As users navigate through a page and consume the content, \systemname implicitly keeps track of the criteria and aspects that they paid attention to on the page based on dwell time, that is, the amount of time they roughly spent reading a block of content, as specified in \cite{liu_crystalline_2022,chen_marvista_2022}.
\systemname then uses this information to summarize the user's research progress in the aforementioned summary block (Figure \ref{fig:selenite-sidebar}f), which consists of three sections: 
(1) criteria that users \textbf{cared about} and have seen evidence for based on implicit tracking (emphasizing their focus and priorities), 
(2) the remaining ones that are discussed on the page but users \textbf{ignored or skipped}\footnote{The threshold of determining if the user indeed paid attention to a paragraph is set to 2 seconds based on our empirical testing. Future work can investigate more adaptive methods, such as taking into account the length of a paragraph, the amount of new information contained in a paragraph compared to users' existing knowledge, or if users appear to be idling and performing irrelevant activities.} (reminding users of potentially overlooked criteria), and 
(3) a set of \textbf{recommended} criteria from the global overview that haven't been discussed in any of the past articles but could still be worth exploring (encouraging users to find additional information about these unseen criteria, for example, by conducting additional searches on them, thereby broadening their perspectives and maximizing information gain).
}

\revision{Specifically, to achieve this third objective, we leverage users' the subset of criteria that users \emph{cared about} 
and the subset they have intentionally \emph{skipped} 
to \emph{recommend} additional \textbf{relevant} and \textbf{diverse} criteria to search for and read about from the remaining global list.} This requirement for the suggested criteria to be both relevant and diverse is similar to the exploration-exploitation trade-off in information retrieval and recommender system literature \cite{athukorala_beyond_2016}, helping maintain user engagement and interest while avoiding over-fit or filter bubbles \cite{kunaver_diversity_2017,sa_diversity_2022}.

\revision{
To operationalize this idea, we consider the problem a graph problem (following the approach by \cite{yang_beyond_2023})}:
by constructing a fully connected graph using the global list of criteria as \emph{vertices}, we assign \emph{edge} weights as distances between respective criteria in a semantic embedding space and vertex weights as the criterion's relevance to the subset of criteria that the user cared about. Our objective then is to recommend a \textit{diverse} subset of criteria (vertices) that have large distances between each other while still being \emph{relevant} to what users cared about.
That is, we need to find a sub-graph $G'$ of size $n$, which maximizes a weighted ($\beta > 0$) sum of vertex weights $w_V$ (relevance) and edge weights $w_E$ (diversity):
%
\[\arg\max_{G' \subset G, |G'|=n} \beta\cdot w_{V}(G') + w_{E}(G') \]

To build the graph, we measure relevance with the perplexity score of the sentence ``\texttt{\{global\_criterion\} tend to be considered together (or is a trade-off) with \{cared\_abou\\t\_criterion\}}'' using \gptTwo~\cite{radford_language_2019}
and characterize diversity with the cosine distance between the SentenceBERT~\cite{reimers_sentence-bert_2019} embeddings of the two vertices (criteria). 
\revision{
Here, we follow the classic greedy peeling algorithm~\cite{wormald_differential_1995} by dropping vertices with the lowest weights (the sum of vertex and every edge weight) one at a time in a greedy fashion until the graph size reaches $n$ (empirically determined as 2). We then present these additional criteria as the ones that users ``might be interested in further searching for'' (Figure \ref{fig:selenite-sidebar}f) in the summary block.}


\subsection{Implementation Notes}

The \systemname browser extension is implemented in HTML, TypeScript, and CSS and uses the React \javascript library \cite{facebook_react_2018} for building UI components. It uses Google Firebase for backend functions, database, and user authentication. 

As explained previously, we implemented \systemname using state-of-the-art NLP models: off-the-shelf \gptFour~\cite{openai_gpt-4_2023} and NLI models finetuned on \texttt{BART}~\cite{lewis_bart_2019}.
These models were chosen for their strong performance \revision{and efficiency} that would satisfy our prototyping needs as well as their generalizability across different application domains (c.f. Section \ref{sec:selenite-performance-eval}). However, it is important to note that \revision{\textbf{our contributions lie more in the concept of grounded reading, interface design, and underlying NLP task abstractions, which are independent of specific model usage}}. We anticipate that these designs will remain valid as AI techniques continue to advance \cite{liu_what_2023}.\footnote{Additional implementation details can be found in section \ref{sec:addition-implementation-details} of the Appendix.}

\section{Study 1: Intrinsic Evaluation}\label{sec:selenite-performance-eval}

While \systemname can help ground users in what to read, its impact may backfire if the list of options and criteria is not accurate or comprehensive --- anchoring bias \cite{tversky_judgment_1974}\footnote{Anchoring bias refers to people's inclination towards relying too excessively on the initial set of information they were exposed to on a topic. Regardless of the accuracy or quality of that information, people use it as a reference point, or an ``anchor,'' to make subsequent judgments or decisions \cite{zong_experimental_2022}.} may cause readers to more easily miss information that is \emph{indeed included in the page} but \emph{not} reflected in the \systemname-generated options and criteria list.
Here we evaluate on \revision{a diverse set of topics} whether \systemname can:
\revision{1) \textbf{accurately} report options that are present on a web page; 2) \textbf{comprehensively} report critical criteria people commonly consider.}


\def\arraystretch{1}
\begin{table*}[t]
\centering
\vspace{-2mm}
\resizebox{1\textwidth}{!}{
\begin{tabular}{r|| cc | ccc || cc | ccc }
\toprule
\multirow{2}{*}{\textbf{Topic}} &
\multicolumn{5}{c}{\textbf{Topic-level}} & 
\multicolumn{5}{c}{\textbf{Paragraph-level}}\\
 \cmidrule(lr){2-6} 
 \cmidrule(lr){7-11}

 & 
 \#GT (total) & \#\systemname (total) & Precision & Recall & F1 &
 \revision{\#GT (avg)} & \revision{\#\systemname (avg)} & Precision & Recall & F1 \\
 \midrule
 \midrule

Best washing machines 
    & 19 & 24 & 0.88 & 1.0 & 0.93 & \revision{3.05} & \revision{3.30} & 0.91 & 1.0 & 0.95 \\
Birthday gift ideas 
    & 11 & 21 & 0.57 & 0.91 & 0.70 & \revision{1.40} & \revision{3.45} & 0.57 & 0.96 & 0.72 \\
Best hybrid app frameworks 
    & 15 & 21 & 0.86 & 0.93 & 0.89 & \revision{2.55} & \revision{3.00} & 0.83 & 1.0 & 0.91 \\
Best time tracking tools 
    & 20 & 21 & 0.81 & 0.95 & 0.87 & \revision{2.95} & \revision{3.20} & 0.88 & 0.98 & 0.93 \\
Deep learning frameworks 
    & 25 & 20 & 0.80 & 0.84 & 0.82 & \revision{3.15} & \revision{3.05} & 0.87 & 0.95 & 0.91 \\
Best sleeping bags 
    & 19 & 21 & 0.81 & 0.89 & 0.85 & \revision{2.85} & \revision{3.15} & 0.95 & 1.0 & 0.97 \\
Best air purifiers 
    & 20 & 24 & 0.83 & 1.0 & 0.91 & \revision{3.05} & \revision{3.75} & 0.83 & 0.98 & 0.90 \\
Best robot vacuums 
    & 23 & 28 & 0.82 & 1.0 & 0.90 & \revision{3.10} & \revision{4.05} & 0.95 & 1.0 & 0.97  \\
Best baby strollers 
    & 22 & 24 & 0.92 & 1.0 & 0.96 & \revision{3.45} & \revision{3.65} & 0.81 & 1.0 & 0.90 \\
Best tropical vacation spots 
    & 15 & 19 & 0.74 & 0.93 & 0.82 & \revision{2.55} & \revision{3.10} & 0.92 & 1.0 & 0.96 \\\midrule
\textbf{Mean} 
    & 19.0 & 22.3 & 0.80 & 0.95 & 0.87 & \revision{2.81} & \revision{3.37} & 0.85 & 0.98 & 0.91 \\
\bottomrule
\end{tabular}
}
\vspace{0mm}
\caption{Statistics of the accuracy and coverage evaluation on \systemname's capability to retrieve a high-quality set of commonly considered criteria by topic. \revision{For topic-level, we report the number of groundtruth criteria, i.e., ``\#GT (total)'', the number of criteria retrieved by \systemname, i.e., ``\#\systemname (total)'', as well as the precision, recall and F1-score. For paragraph-level (recall that we randomly sampled 20 paragraphs per topic), we report the average number of groundtruth criteria mentioned per paragraph, i.e., ``\#GT (avg)'', the average number of criteria reported by \systemname, i.e., ``\#\systemname (avg)'', as well as the precision, recall and F1-score.}}
\label{tab:performance-eval}
\vspace{-5mm}
\end{table*}

\subsection{Methodology}\label{sec:selenite-performance-eval-methodology}

\subsubsection{Topic Sampling}
We collected ten topics that exhibit a mixture of practicality and diversity (Table \ref{tab:performance-eval}): (1) we randomly sampled five topics (out of the 28 topics reported) reported by participants in the formative study, and (2) we collected five more from Wirecutter, a popular review site --- the three most popular product guides listed in their 2021 year-in-review (at the time of writing, the 2022 year-in-review has not been published) as well as their two most recently updated guides for June 2023. 

\subsubsection{Groundtruth Dataset Creation for Options and Criteria}
To collect groundtruth criteria that the general audience would care about for each topic, we mimic a typical information collection workflow, where people rely on top sources from popular search engines for their authenticity and credibility. Specifically, we first gathered the top five Google search results using the query template ``\texttt{best [product or category]}'' (excluding promotions or ads). \revision{We show a partial snapshot of a representative web page in Figure \ref{fig:selenite-example-webpage} in section \ref{sec:apx-example-webpage} of the Appendix.}
Then, for each web page, two authors independently first read through and annotated the options and criteria mentioned in every paragraph, and then merged all the annotations, excluding duplicate ones. Note that since many criteria are mentioned in a descriptive manner (e.g., the phrase ``It is available in a black finish'' implicitly refers to \criterionNameSmall{aesthetics}), the two authors had some variance in how they named essentially the same criteria. Therefore, the two authors iteratively discussed and resolved their conflicts, merging criteria that they believed were semantically equivalent, producing the groundtruth criteria list \revision{(see Table \ref{tab:selenite-groundtruth-criteria} in section \ref{sec:apx-groundtruth-criteria} of the Appendix for the groudtruth criteria list)}.
Finally, for the topics that we sampled from the formative studies where participants explicitly collected options and criteria using \unakite, we double-checked and were able to verify that all the criteria that they identified were indeed included in our groundtruth dataset, providing preliminary evidence to the soundness of our groundtruth dataset.\looseness=-1


\subsubsection{Evaluation Metrics}

\paragraph{Option Extraction} Since \systemname directly extracts options from web pages, we evaluated this capability using the \emph{accuracy}, that is, the percentage of options extracted by \systemname out of all the options available on a page.

\paragraph{Criteria Retrieval} We also evaluated \systemname's ability to retrieve the right set of criteria on two levels. First, to answer whether \systemname helps find useful criteria \emph{for each topic}, \revision{we compute \emph{topic-level precision} (``the fraction of criteria retrieved by \systemname that coincided with the groundtruth'') and \emph{recall} (``the fraction of groundtruth criteria that are were also retrieved by \systemname'').}

Second, to measure whether \systemname provides high-quality groundings \emph{per paragraph}, we additionally randomly sampled 20 paragraphs per topic, and \revision{computed \emph{paragraph-level precision} (``the fraction of criteria recognized by \systemname that were indeed mentioned in the paragraph'') and \emph{recall} (``the fraction of criteria mentioned in the paragraph that were indeed reported by \systemname'').}

\subsection{Results}

\subsubsection{Option Extraction}
\systemname achieved 100\% accuracy on extracting options from web pages, i.e., as long as there was an option explicitly mentioned on a web page, \systemname was able to correctly extract it. This directly speaks to the strong reasoning and information extraction capabilities of \gptFour as described in OpenAI's technical report \cite{openai_gpt-4_2023}.

\subsubsection{Criteria Retrieval}
We present the result of criteria retrieval evaluation metrics in Table~\ref{tab:performance-eval}, which provides initial evidence to \systemname's strong capability in presenting to the user a comprehensive set of criteria that people commonly consider.
Notice that for most of the topics, \systemname retrieved \textit{more} criteria compared to the groundtruth set. This is not surprising, partly due to the fact that \gptFour has likely synthesized information from significantly more sources than what was considered during the construction of the groundtruth dataset (five web pages for each topic). Theoretically, there is also a possibility that \gptFour hallucinated some criteria that are largely irrelevant to a given topic, however, upon further manual inspection, we did not see evidence of hallucination, at least for the 10 topics considered in this evaluation (for example, Table \ref{tab:selenite-baby-stroller-sample-criteria} shows a list of commonly considered criteria that \systemname retrieves for the topic of ``best baby strollers'').

\paragraph{Topic-level recommendations} 
\systemname achieved both high recall and high precision on multiple topics (e.g., \emph{best washing machines}, \emph{best air purifiers}, \emph{best robot vacuums}, and \emph{best baby strollers}), and usually achieves higher recall than precision, suggesting that \systemname has the tendency of finding \emph{supersets} of what users would generally be able to identify from reading, i.e.,  criteria in the groundtruth set.\looseness=-1

We qualitatively analyzed the topics with a lower-than-average topic-level criteria recall, and found two contributing reasons: 
(1) Some web pages cover factual information that is not necessarily relevant.
Multiple pages describing \emph{Best Hybrid App Frameworks} mentioned ``First Release Date,'' which arguably is not a \emph{criterion} necessary for selection.
(2) Some criteria are inter-correlated. 
For example, in the case of \emph{deep learning framework}, whereas it did not explicitly mention ``\texttt{growth speed},'' \systemname did suggest \criterionNameSmall{innovation}, whose description is ``{\small\texttt{the ability of the framework to stay up-to-date with the latest research and developments in deep learning, and to incorporate new techniques and architectures as they emerge.}}''
While we did not count these two as equivalent in the evaluation, in practice, these two have a high correlation, and we believe having one included might be sufficient. 
Still, this potential mismatch reflects the necessity of allowing users to edit the criteria and descriptions.  
Meanwhile, upon initial observation, \systemname's lower precision on certain topics may suggest its inclination towards retrieving unnecessary criteria. However, a closer examination revealed an interesting insight: 
for instance, when it comes to topics like \emph{birthday gift ideas}, popular web pages often present a list of 10+ diverse options that lack strict comparability and are all described using generic terms such as ``fun'' or ``sweet.'' This lack of specificity makes it challenging to determine a comprehensive set of groundtruth criteria. In contrast, \systemname offers comprehensive overviews that encompass factors like \criterionNameSmall{personalization}, \criterionNameSmall{uniqueness}, \criterionNameSmall{practicality}, \criterionNameSmall{sentimentality}, and \criterionNameSmall{presentation (wrapping)}, among others.\looseness=-1 

\paragraph{Paragraph-level Grounding}
\systemname also achieved high per-parag-raph performances, again with a bias towards higher recalls.
This is intentional --- we tuned the parameters of the NLI-based method such that it is more likely for \systemname to claim non-existing criteria than overlooking actual existing ones. This approach prioritizes avoiding information loss, which, suggested by prior work \cite{liu_what_2023}, is a more expensive mistake compared to user verification.
We order the criteria based on their probability score from the NLI model and will, in future iterations, fade the ones with a lower score.\looseness=-1

We did notice that in some rare scenarios, the NLI performance can be influenced by a criterion's description, e.g., changing ``appropriate for age'' to ``appropriate for kids, adults, or elderly'' can reduce \systemname's error on recognizing arbitrary numbers as ages. 
Therefore, in future iterations of \systemname, we will provide a hint to users, prompting them to try tweaking the description when they attempt to delete a criterion due to its seemingly low grounding efficacy.\looseness=-1

\section{Study 2: Usability Evaluation}\label{sec:selenite-study-2-usability}

\def\arraystretch{1}
\begin{table*}[t]
\vspace{-2mm}
\centering
\resizebox{0.9\textwidth}{!}{%
\begin{tabular}{r|l|l|l|l|l|l}
    \toprule
    ~ & \textbf{Mental demand} & \textbf{Physical demand} & \textbf{Temporal demand} & \textbf{Performance} & \textbf{Effort} & \textbf{Frustration} \\
    \midrule
    \textbf{\systemname}
    & $3.0~(3.03 \pm 1.76)$* 
    & $1.0~(0.51 \pm 1.74)$
    & $2.5~(2.26 \pm 1.68)$*
    & $8.5~(8.47 \pm 1.32)$*
    & $3.5~(4.08 \pm 1.88)$*
    & $0.5~(0.33 \pm 1.51)$
    \\
    \midrule
    \textbf{Baseline}
    & $6.5~(6.43 \pm 2.07)$*
    & $1.0~(0.79 \pm 1.98)$
    & $4.0~(4.29 \pm 2.08)$*
    & $6.5~(6.54 \pm 1.73)$*
    & $6.0~(5.98 \pm 2.23)$*
    & $1.0~(0.89 \pm 1.91)$
    \\
    \bottomrule
\end{tabular}%
}
\caption{Study 2 participants' responses to NASA TLX questions (on a scale from 0 to 10) in study 2. Format: median (mean $\pm$ standard deviation). Statistically significant differences (p < 0.05) through t-tests are marked with an *.}
\label{tab:selenite-tlx}
\vspace{-5mm}
\end{table*}

\begin{table*}[t]
\centering
\resizebox{0.9\textwidth}{!}{
\begin{tabular}{
>{\raggedright}r|
p{114mm}|
p{20mm}
}
\toprule
\textbf{Question category} & \textbf{Statement} &
\textbf{Response}
\\\midrule

\textbf{Comprehensibility} & I would consider my interactions with the tool to be understandable and clear. &
$6~(6.33 \pm 1.10)$ 
\\\midrule

\textbf{Learnability} & I would consider it easy for me to learn how to use this tool. &
$7~(6.71 \pm 1.04)$ 
\\\midrule

\textbf{Enjoyability} & I enjoyed the features provided by the tool. &
$6~(6.13 \pm 1.72)$ 
\\\midrule

\textbf{Applicability} & Using this tool would make solving sensemaking problems more efficient and effective. &
$6~(6.28 \pm 1.39)$ 
\\\midrule

\textbf{Recommendability} & If possible, I would recommend the tool to my friends and colleagues. &
$6~(6.23 \pm 0.94)$ 
\\\bottomrule
\end{tabular}
}
\caption{Study 2 participants' responses to System Usability Scale questions (on a scale of 1 to 7, where 1 represents ``strongly disagree'' and 7 represents ``strongly agree'') in study 2 regarding their \systemname experience. Format: median (mean $\pm$ standard deviation)}
\label{tab:selenite-sus}
\vspace{-5mm}
\end{table*}

We also conducted an initial usability study to evaluate if the features provided by \systemname are usable and if the approach of providing global as well as contextual grounding can allow users to read, navigate, and comprehend information more efficiently. 
Specifically, we were interested in the following quantitative research questions:

\begin{itemize}[leftmargin=.2in]
    \item \textbf{[RQ1]} Does using \systemname speed up people's process of reading and understanding information?
    \item \textbf{[RQ2]} Does using \systemname help people achieve a more comprehensive understanding of an information space?
    \item \textbf{[RQ3]} Can \systemname help people obtain new information in addition to their existing knowledge?
\end{itemize}

\subsection{Methodology}
We recruited 12 participants (five female, seven male) aged 21-40 ($\mu$ = 28.9, $\sigma$= 5.2) through social media. Participants were required to be 18 or older and fluent in English. 
All participants reported that they regularly engage in the process of seeking and sifting through large volumes of online information, whether for professional or personal purposes, on a weekly basis.\looseness=-1

The study was a within-subjects design, where participants were presented with two tasks and were asked to complete each one under a different condition, counterbalanced for order. \revision{For each task, participants were given a topic that they needed to investigate and two web pages relevant to the topic that they were required to read and process. The two topics were \emph{``best baby strollers''}
and \emph{``best robot vacuums.''}\footnote{To ensure realism and participant engagement, the tasks were selected based on actual topics that the formative study participants reported investigating. Rather than letting participants search for their own pages to read from the get-go, we provided them with a predefined set of pages to enable a fair comparison of the results (e.g., speed, etc.). 
Requiring participants to use predefined pages (each contains, on average, 15 screenfuls of content) for the first portion of the study also helps ensure that the two tasks are of roughly equal difficulty in terms of reading and cognitive processing effort. As described in the results, there was no significant difference by task.}
}
\revision{The provided two web pages for each topic were all product comparison pages used in the previous study (see section \ref{sec:selenite-performance-eval-methodology}).} 
For each task, participants were 
asked to read through the two required pages, either by themselves without any aid (a \emph{control} condition simulating how people normally read) or with \systemname (\emph{experimental} condition). While reading, they were instructed to write down \emph{as many} criteria as they learned and thought were important for the topic as well as the reason why they were important as if they needed to thoroughly explain the topic to a friend later. \revision{Then, participants were instructed to optionally search (using Google) and gather additional information that they still wanted to learn about but weren't able to from reading the two required pages.}
We imposed a 25-minute limit per task to keep participants from getting caught up in one of the tasks. However, they were instructed to inform the researcher that they felt like they could make no further progress, i.e., having learned as much as they could about the given topic.

Each study session started by obtaining consent and having participants fill out a demographic survey. Participants were then given a 5-minute tutorial showcasing the various features of \systemname and a 5-minute practice session before starting. At the end of the study, the researcher conducted a NASA TLX survey and a questionnaire, eliciting feedback on their experience in both conditions. 
Each study session took around one hour, using a designated Macbook Pro computer with the latest version of Chrome and \systemname installed, and was conducted remotely via Zoom. 
Each participant was compensated with \$15 USD. The study was approved by our institution's IRB.\looseness=1

\subsection{Results}

All participants were able to complete all tasks in both conditions, and nobody went over the pre-imposed time limit. Below, we present primarily quantitative evidence to evaluate the usability 
of \systemname with respect to our research questions.\looseness=-1

First, we were interested in understanding if \systemname can help participants read and process information faster compared to the baseline condition (RQ1). To examine this, we measured the time it took for them to finish reading all the materials in each task. A two-way repeated measures ANOVA was conducted to examine the within-subject effects of the condition (baseline vs. \systemname) and task on completion time. There was a statistically significant effect of condition (F(1, 20) = 102.5, p < 0.01) such that participants completed tasks significantly faster (36.3\%) with \systemname (Mean = 840.3 seconds, SD = 102.7 seconds) than in the baseline condition (Mean = 1319.3 seconds, SD = 120.0 seconds). There was no significant effect of task (F(1, 20) = 0.40, p = 0.53), indicating the two tasks were indeed of roughly equal difficulty. These results suggest that \systemname helped participants read and comprehend information more efficiently. 
We discuss additional qualitative insights into why \systemname was more efficient in the following open-ended case study (section \ref{sec:selenite-open-ended-study-qual-findings}).

In addition, we were interested in understanding if \systemname can help participants achieve a more comprehensive understanding of a topic (RQ2). To measure this, we first compared the \emph{quantity} of criteria that participants externalized under each condition. As a pre-filtering step, two researchers rated all the criteria that participants externalized as either \emph{valid} or \emph{invalid} blind to the conditions. Valid criteria are considered as ones that are \emph{relevant} to the topic and \emph{backed by specific evidence} that can be traced back to the content, consistent with those standards used by prior work in judging the quality of subjective evidence \cite{chang_mesh_2020}. After resolving conflicts (which were minimal) between the two researchers and filtering out the criteria that were invalid, we found that the average total number of valid criteria increased by 90.4\% when using \systemname (Mean = 12.93, SD = 3.90) compared to the baseline condition (Mean = 6.79, SD = 4.07), which is statistically significant (p < 0.01) under a t-test. Thus, using \systemname appeared to enable participants to identify and learn significantly more criteria about a topic compared to people's current way of reading information.

\vspace{2mm}
In addition to quantity, we also examined the \emph{quality} of the criteria by comparing the ones that participants externalized with the groundtruth criteria curated in the previous accuracy and coverage evaluation --- we can calculate the precision (calculated as $n_{\textbf{Hit}}/n_{\textbf{Total}}$) and recall (calculated as $n_{\textbf{Hit}}/n_{\textbf{Groundtruth}}$) of participants' criteria that \emph{hit} the groundtruth (where $n_{\textbf{Total}}$ is the total number of valid criteria participants externalized, and $n_{\textbf{Groundtruth}}$ is the number of groundtruth criteria for each task). On average, \revision{participants in the \systemname condition achieved significantly higher precision (98.8\% vs. 78.4\%, p < 0.05) and recall (73.0\% vs. 30.4\%, p < 0.05) in both tasks.} Thus, using \systemname appeared to have enabled participants to improve the quality of their understanding of an information space in terms of its criteria.\looseness=-1

Furthermore, to understand if \systemname can help participants obtain new information in addition to what they have already learned from reading the two required pages (RQ3), we examined: 
1) the number of additional \textit{searches} that they performed in the \systemname condition (Mean = 2.01, SD = 1.39), which turned out to be significantly more (p < 0.05) than the baseline condition (Mean = 0.33, SD = 0.62); 
2) the number of additional \textit{pages visited} in the \systemname condition (Mean = 2.76, SD = 2.32), which turned out to be significantly more (p < 0.05) than the baseline condition (Mean = 0.42, SD = 0.74); and 
3) the number of additional \textit{criteria} that participants externalized in the \systemname condition (Mean = 1.58, SD = 0.91), which turned out to be significantly more (p < 0.05) than the baseline condition (Mean = 0.33, SD = 0.22). These results suggest that \systemname did encourage and help participants to seek additional information beyond their existing perspective.\looseness=-1

Last but not least, participants filled out a NASA TLX \cite{hart_development_1988} cognitive load scale and a System Usability Scale (SUS) \cite{lewis_system_2018} questionnaire for each condition. SUS Likert items were integer-coded on a scale from 1 (strongly disagree) to 7 (strongly agree). The median response values are presented in Tables~\ref{tab:selenite-tlx} and~\ref{tab:selenite-sus}. Notably, participants perceived \systemname to have significantly lowered workload across mental, temporal, and effort demands as well as significantly increased perceived performance based on paired t-tests). This suggests that using \systemname can reduce the cognitive load and interaction costs when reading and understanding information, even when users had to learn and get used to a new user interface.\looseness=-1

\section{Study 3: Open-ended Case Study}\label{sec:selenite-study-3-usefulness}

Encouraged by the promising performance outcome of the previous two studies, we conducted a third open-ended case study to understand the usefulness and effectiveness of the \systemname prototype from a qualitative perspective.

\subsection{Methodology
}
We recruited eight participants (three male, five female; three students, two software engineers, one dermatologist, one accountant, and one researcher) aged 24-55 years old (Mean = 33.6, SD = 8.1) through emails and social media. The same recruitment requirements were applied, but individuals who participated in the previous usability study were excluded from this study.\looseness=-1 

\revision{Each participant first completed two pre-defined tasks, where they used \systemname to help them read information about an unfamiliar topic. From the topics that participants reported having explored in the formative study,} we randomly selected two that the participant was unfamiliar with (indicated in their screening survey). For each task, participants were presented with a set of three web pages that covered the topic that the formative study participants had gone through. The provided web pages were primarily review articles comparing several options together or product detail pages.\looseness=-1
We imposed a 20-minute limit per task to keep participants from getting caught up in one of the tasks.
\revision{
To further explore \systemname's potential, all participants then used \systemname to make sense of a third topic that they intend to explore in real-life. 
Here, we purposefully did not limit the topic to be ``unfamiliar,'' allowing participants to revisit previous topics of interest and potentially uncover fresh perspectives.
Each study session began by obtaining consent and demographic information.} 
Participants were then given a 5-minute tutorial showcasing the various features of \systemname and a 5-minute practice session before starting. 
\revision{At the end of the study, the researcher elicited feedback on using \systemname through a semi-structured interview, which was recorded and later transcribed for coding and thematic analysis \cite{charmaz_constructing_2006}.}
Each study was conducted via Zoom for up to one hour. Each participant was rewarded with \$15 USD. The study was approved by our institution's IRB.

\subsection{Results}\label{sec:selenite-open-ended-study-qual-findings}

Below, we present the major qualitative findings from the observation of participants' behaviors using \systemname as well as their feedback from the post-study interviews.\footnote{To see all the topics that participants explored in this study, please refer to Table \ref{tab:selenite-study-3-topics} in the Appendix.}

\paragraph{Time and effort savings.} All of the participants mentioned that using \systemname would save them a lot of time and effort compared to using their typical reading and information collection workflow, echoing the quantitative results reported in the usability evaluation (see section \ref{sec:selenite-study-2-usability}). First of all, having access to the global overview felt like \userquote{a game-changer} (P8) that offers a \userquote{bird's-eye view} (P4) or access to \userquote{on-demand expert opinion} (P1) that \userquote{took away the anxiety and guesswork of wondering what other folks would actually care about} (P5). P7 suggested that \userquote{this is something that I always wished for when reading about stuff that I'm not an expert in. It seriously saves me a ton of time that I'd otherwise spend trying to wrap my head around it little by little,} while P6, who couldn't \userquote{stand the huge deal of work of figuring out stuff that I'm not used to} said \userquote{now I really feel like I'm chilling in the passenger seat and not having to do all the heavy-lifting personally.}\looseness=-1

Second, participants seemed to appreciate the in-context annotations and summaries of each paragraph provided by \systemname. They thought that this feature \userquote{made things incredibly easy} (P3) by \userquote{helping me grasp the key points without wasting time reading a paragraph through} (P1), and \userquote{felt like back in the day when my classmate would mark all the important stuff in the textbook after a class when I couldn't make it.} However, some did report that the in-context annotations can occasionally be \userquote{a little bit distracting}, especially for paragraphs that are \userquote{apparently unrelated to the main content} (P2), such as those that talk about related articles or terms of services, suggesting that future versions of \systemname should consider more robust content filtering techniques.

Last but not least, participants also appreciated that \systemname can help them brainstorm search queries that would enable them to find new information more efficiently that was \userquote{almost always one step ahead} (P4), especially in the third task. For example, after reading two review articles about e-readers, \systemname suggested that P5 could do some additional investigations about \criterionNameSmall{supported file formats} and \criterionNameSmall{syncing across devices}. P5 admitted that \userquote{I'd totally miss those if I'm by myself, and even if I'm trying to be super careful, it would take me forever to figure out that I need to check out those aspects.} Additionally, we observed that when integrating the \systemname suggested criteria into subsequent search queries, the search engine did return result pages that turned out to be noticeably different yet sufficiently high-quality for users to explore.\looseness=-1

\paragraph{Impact on reading patterns and habits.}
Participants all mentioned that they immediately checked out the commonly considered criteria from the sidebar before diving into reading the first web page. They claimed that compared to what they normally do, which is \userquote{just have to hunker down and read}, reading the overview first helped them \userquote{cut to the chase and get a feel of what's out there} (P7) and remind them of criteria that would otherwise \userquote{slip my [their] mind} (P3).\looseness=-1

On a per-paragraph level, we noticed an initial hesitation among some participants (3 out of 8) towards relying solely on the provided criteria labels. As a safety precaution, they personally read through a handful of paragraphs to confirm the labels' accuracy and reliability. We further corroborated this observation with their reflections, such as \userquote{I've never seen anything like this before, so honestly, I was a bit skeptical at first. But hey, everything looked legit!} (P5) After this initial hurdle, participants tend to \userquote{rely on the labels to tell me the gist of a paragraph} (P4) and only read paragraphs that discuss criteria that they truly cared about. For the content that participants did end up reading, they think the corresponding criteria \userquote{definitely helped me [them] process and digest it better} (P6), and even \userquote{saved me [them] from otherwise misunderstanding things} (P7). For example, while exploring healthy diet plans, P7 reflected that he would have initially thought a paragraph detailing the caloric allocation for each meal was seemingly discussing ``calorie intake,'' however, \systemname preemptively clarified that the focus was on \criterionNameSmall{portion control}, i.e., ``{\small\texttt{providing guidelines on portion sizes.}}''\looseness=-1

Participants also enjoyed the easy navigation feature that \systemname offers, and used it to frequently jump between different criteria mentionings for easy comparison and digestion (7/8). They claimed that finding specific criteria about different options in a long article used to be \userquote{link finding a needle in a haystack} (P8) that they were hesitant to do, but with \systemname, \userquote{it's more like following a well-lit path} (P5). For example, P2 reflected on her experience exploring VPN solutions, and claimed that \userquote{now I get it, McAfee Safe Connect seems to be keeping track of all sorts of my information while SurfShark doesn't do any of that. If I can't quickly switch between these two points on the page, by the time I reach SurfShark's no-logging policy, I would have totally forgotten about what McAfee does, or that I should even be concerned about logging at all.} In addition, participants liked the fact that they can more effectively break out from the original structure and narrative of an article; for instance, P1 recounted that \userquote{you don't gotta stick to what the authors say anymore, ya know? Because, let's face it, their storylines can get all tangled and complicated sometimes.}

Last but not least, we did not observe much usage of the ``zoom in'' feature, where \systemname can leverage \gptFour to provide a thorough analysis of a piece of content --- only 3 participants tried it for a total of 8 times. We hypothesize that 1) the web pages utilized in the study were all professionally crafted, resulting in content that was relatively easy to comprehend; 2) the criteria labels generated by our NLI pipeline proved to be adequate in addressing the participants' information needs; 3) the time required for the ``zoom in'' feature to provide a useful analysis, typically ranging from 5 to 10 seconds, still exceeded the participants' patience and attention span. Future work could explore solutions to address this limited adoption from these perspectives, for example, with models that boast significantly increased inference speeds.\looseness=-1

\vspace{-2mm}

\paragraph{Additional findings.}

One interesting theme that emerged was that some participants (4/8) opted to use \systemname in the third task to revisit topics that they had previously explored and wanted to be able to \userquote{double-check} (P3) whether their prior understanding of the topic was truly comprehensive. Consistently, each participant uncovered something new that they hadn't considered before. For example, P3, who had recently been making plans to move in with his partner, revisited the topic of ``\emph{choosing the right mattress},'' and realized that he had never taken into consideration criteria such as \criterionNameSmall{motion transfer} (i.e., ``{\small\texttt{the extent to which movement on one side of the mattress affects the other side}}'') or \criterionNameSmall{noise reduction} (i.e., ``{\small\texttt{the ability of the mattress to minimize noise from springs, coils, or other components}}''), which prompted him to reassess his original mattress purchase. As another example, P4, a professional software engineer, revisited the topic of ``\emph{choosing a hybrid app framework}'' and discovered that he had neglected to consider the \criterionNameSmall{licensing and legal considerations} (i.e., ``{\small\texttt{compliance with licensing requirements and legal considerations}}'') as suggested by \systemname. Consequently, P4 was able to find additional evidence to confirm the validity of their original framework choice made back in 2017.\looseness=-1

In the post-study interview, many participants (6/8) felt that now they \userquote{can't imagine reading without a tool like this (\systemname)} (P3). Half even inquired about the possibility of installing \systemname on their personal computers for post-study usage, and we gladly fulfilled their requests. Despite encountering a few bugs in our research prototype during the study and having no obligation or incentives for continued usage after the study, the fact that they were willing to do so suggests that our grounded reading approach indeed holds value for our participants.

\section{Discussion}

Some of the participants (3/8) from the case study expressed concern about the coverage of \systemname's overview criteria and the criteria labels for each paragraph at the beginning. They wondered if \systemname might overlook important criteria that they should also consider. This concern was valid, given that we presented the tool as an AI-powered oracle that could potentially be fallible or overlook certain factors and encouraged users to conduct their own explorations in addition to relying on \systemname's insights. However, our accuracy and coverage evaluation described in section \ref{sec:selenite-performance-eval} provides an initial validation that the criteria and options provided by \systemname are indeed comprehensive, relevant, and accurate. In addition, after the study, participants also acknowledged that the current set of criteria offered by \systemname already \userquote{far exceeds what I [they] could identify and keep track of on my [their] own} (P4); therefore, they \userquote{wouldn't mind at all if the algorithm misses any minor ones} (P1). Indeed, despite the participants' awareness of the opportunity to request additional criteria from \systemname in the case of insufficient coverage (as confirmed in the post-study interviews), we did not observe any instances of such usage. \revision{Nevertheless, further research is necessary to investigate: 1) ways that would further improve the coverage and accuracy of \systemname, such as leveraging retrieval-augmented models \cite{lewis_retrieval-augmented_2020}; 2) mechanisms and interventions designed to reduce over-dependence on \systemname as well as encourage critical thinking and  user-led explorations.}\looseness=-1

Though primarily designed as a tool for grounded reading, \systemname might also have the potential to address some of the issues identified by prior work regarding structuring information during sensemaking --- prior research suggested that asking users to structure information too early might lead to a more poorly structured information space \cite{kittur_costs_2013}. In addition, the knowledge structures that people created often become obsolete, and new structures often emerge as their mental representations evolve over the course of their investigation \cite{fisher_distributed_2012,kittur_costs_2013,hearst_whats_2014}, resulting in having to spend significant effort in refactoring the structures every once in a while. Here, \systemname provides users with a well-structured framework from the outset, including a set of commonly acknowledged criteria. This readily usable scaffold serves as a starting point, aiming to encompass the majority's perspective and thereby minimizing the necessity of refactoring or restructuring. Hopefully, it simplifies the iterative and cognitive-demanding process of building a mental model, transforming it into a possibly more manageable task of refining and pruning \cite{liu_crystalline_2022}.\looseness=-1 

There was also a concern that \gptFour could potentially hallucinate or generate irrelevant or even false criteria and thus mislead users in their subsequent exploration. However, it is important to note that in our case study, as well as in study 2, we did not observe any such episodes or evidence of this occurring. This could be attributed to the fact that the topics explored in the study were all common subjects with abundant source materials available online, which were likely encountered by \gptFour during its training process. We would like to further conjecture that even if hallucination occurs, users can readily identify irrelevant or false criteria by carefully reading their descriptions and comparing them with common sense or their intuitive knowledge about the topic, mitigating the actual impact of hallucination.

\section{Limitations \& Future Work}

\paragraph{Connections among criteria.}
In \systemname, we made the implicit assumption that criteria are completely independent. However, in reality, there could be connections between the criteria
--- for instance, when evaluating the ``Best Baby Stroller,'' the specific criterion of \criterionNameSmall{suspension system} falls under the broader category of \criterionNameSmall{safety} (\emph{hierarchy}), while on the other hand, aspects like \criterionNameSmall{price} and \criterionNameSmall{versatility} are typically trade-offs that are impractical to optimize for simultaneously (\emph{correlation}). 
Currently, \systemname takes into account one form of criteria connections, i.e., \emph{relevance} between criteria, when suggesting the next steps. This proved promising in the study, which gave us reasons to believe that further exploiting these connections between criteria can better support users' reading. For example, instead of presenting the criteria in a simple list, one can imagine creating a behind-the-scenes knowledge graph where criteria are connected using edges of relations (\texttt{TypeOf}, \texttt{CompetesWith}, etc.). By initially displaying a portion of this graph and allowing users to ``zoom in'' on the specific criteria they are interested in (e.g., a subset of ``safety''-related features), we can help users intuitively reason through an initially overwhelming list. In addition, one can again imagine ``overview first, details later''-style UIs \cite{shneiderman_designing_2000} that accommodate criteria hierarchies, e.g., multi-level tables or lists, granting users the flexibility to combine or decompose criteria at decision time.\looseness=-1

\paragraph{Availability of domain knowledge in LLMs.} LLMs, such as \gptFour, possess an extensive range of encoded knowledge, yet they might lack domain-specific information for specialized or emerging topics, as well as for topics involving confidential or sensitive information. Our technical implementation in \systemname is primarily based on extracting knowledge (e.g., commonly considered criteria) from commercially available LLMs, and its effectiveness is highly dependent on the LLM's capability to capture and synthesize relevant domain knowledge from its training data. Without such knowledge, the guidance provided may be subpar. In addition, LLMs themselves can sometimes be biased, and the response they generate might be incorrect or harmful \cite{kumar_language_2023,nadeem_stereoset_2021}. However, our approach to ground the reading process with domain knowledge would also work with other sources of knowledge bases as well, for example, \unakite + \strata tables \cite{liu_reuse_2021,liu_unakite_2019}, or crowdsourced \cite{hahn_knowledge_2016,metaxa_auditing_2021}, or a combination of them. Furthermore, we should also urge users to thoroughly examine the \systemname overview when dealing with critical situations.

\paragraph{Generalizing beyond comparison tasks.} In this work, we focus on helping people \revision{with sensemaking tasks that  often need users to systematically compare different options with respect to various criteria}. As evidenced by our formative as well as case studies, it is beneficial for people to be aware of the criteria that other people commonly consider upfront to help with their subsequent sensemaking journey. However, there are other types of sensemaking tasks, such as those that are purely exploratory or investigatory (e.g., debugging, learning a new skill), that do not entail well-established options and criteria, and as such, they are not ideally compatible for \systemname to assist with. Nevertheless, we postulate that the notion of procuring comprehensive expert perspectives upfront may still apply in \revision{these non-comparison tasks} --- for example, one can imagine asking an LLM for advice on a range of typical strategies to try when debugging or a list of common steps to take to master a new skill. Future research can work on enabling users to obtain these categories of overviews by adapting and customizing the LLM prompts used in \systemname (that were originally used to obtain criteria) and continue to receive similar in-context annotations and reading guidance grounded on those overviews.\looseness=-1

\paragraph{Impact on learning.} The current design of \systemname functions as an ``index'' to direct users to relevant parts of a web page for reading and processing. However, we need to be cautious about a potential risk associated with this approach --- some users might believe they have gained sufficient knowledge about a topic by merely reading the overview and may, therefore, skip engaging with the actual web content. This behavior could lead to incomplete, biased, or even inaccurate understandings of the subject. It is akin to only reading the table of contents or indices of a book without delving into the actual passages. Nevertheless, our studies conducted under controlled settings have shown that participants did, in fact, engage with the actual web content after going through the overviews. To build on this promising evidence, future research should additionally investigate interface and interaction designs that motivate users to explore and read the actual web content with the assistance of \systemname-style guidance. One potential approach could be progressively revealing criteria information to users based on their reading behavior, encouraging deeper exploration and understanding.\looseness=-1


\paragraph{Field Study.} In the future, once all the bugs and usability issues have been thoroughly addressed, we aim to conduct an extensive, long-term field study on a larger scale, where people will have both sufficient motivations to investigate topics relevant to their own personal context and familiarity with \systemname through repeated usage. \revision{This could potentially shed light on situations where \systemname performs reasonably well, as well as situations where it may fall short. Additionally, we are also interested in understanding \systemname's long-term impact on individuals' analytical skills and problem-solving abilities.}\looseness=-1

\section{Conclusion}

\revision{Sensemaking in unfamiliar domains can often be challenging,} with users having to sift through large volumes of information and compare different options with respect to various criteria. 
Previous sensemaking research as well as our new formative study has shown that people would benefit from seeing an overview of the information space, such as the criteria that others have previously found useful. 
However, existing systems have been limited by the ``cold start'' issue --- they require substantial effort from previous users to gather and structure information to produce such an overview, and, even if it has been produced, there are no straightforward methods of sharing that overview with future users and making it so that future users would find it to be comprehensive, unbiased, and useful.\looseness=-1

In this work, by leveraging recent advances in LLMs and natural language processing, we introduce a novel system named \systemname that automates finding an initial set of options, criteria, and evidence, and provides a comprehensive overview to users at the start of their sensemaking process. In addition, it also adapts as people use it, helping users find, read, and navigate unfamiliar information in a systematic yet personalized manner. As such, it provides a valuable proof of concept of how a future LLM-powered sensemaking tool that provides users with comprehensive overviews and in-context reading guidance can \revision{scaffold their sensemaking and learning of an unfamiliar space.}\looseness=-1


\begin{acks}
This research was supported in part by NSF grants CCF-1814826 and FW-HTF-RL-1928631, Google, Adobe, Oracle, Bosch, the Office of Naval Research, and the CMU Center for Knowledge Acceleration. 
Any opinions, findings, conclusions, or recommendations expressed in this material are those of the authors and do not necessarily reflect the views of the sponsors. 
We sincerely thank Daniel M. Russell and Kenneth Holstein for their extraordinary feedback and support.
\end{acks}

\bibliographystyle{ACM-Reference-Format}
\bibliography{references}


\clearpage
\newpage
\onecolumn
\appendix

\section{Unfamiliar Topics Participants Explored in the Formative Study}

\begin{table}[h]
\centering
\resizebox{1\textwidth}{!}{
\begin{tabular}{r|l|l|l}
\toprule
\textbf{Index} & \textbf{Theme}                                      & \textbf{Topic}                                                  & \textbf{Participants} \\\midrule\midrule
1              & \multirow{7}{*}{Software \& online services}        & Choosing a hybrid app framework                                 & P2, P5                \\
2              &                                                     & Selecting a secure password manager                             & P3, P7                \\
3              &                                                     & Choosing a suitable ERP (Enterprise Resource Planning) solution & P1                    \\
4              &                                                     & Choosing a reliable VPN (Virtual Private Network) provider      & P1, P5                \\
5              &                                                     & Picking a deep learning framework                               & P8                    \\
6              &                                                     & Deciding on the best data visualization tool                    & P6                    \\
7              &                                                     & Choosing the best time tracking tool                            & P4                    \\\midrule
8              & \multirow{5}{*}{Consumer Electronics \& Technology} & Choosing a high-quality digital camera                          & P2, P5                \\
9              &                                                     & Choosing the best action camera                                 & P8                    \\
10             &                                                     & Selecting a VR headset                                          & P7                    \\
11             &                                                     & Picking a drone                                                 & P3, P5                \\
12             &                                                     & Picking a smart home ecosystem                                  & P1, P6, P8            \\\midrule
13             & \multirow{5}{*}{Home Appliances \& Furniture}       & Picking the best robot vacuum                                   & P2, P5                \\
14             &                                                     & Choosing the best air purifier                                  & P4                    \\
15             &                                                     & Selecting the best washing machine                              & P3                    \\
16             &                                                     & Picking the right refrigerator                                  & P3                    \\
17             &                                                     & Selecting the best mattress                                     & P3, P6                \\\midrule
18             & \multirow{3}{*}{Outdoor \& Adventure}               & Choosing the best city bike                                     & P7                    \\
19             &                                                     & Choosing the best barbecue grill                                & P3                    \\
20             &                                                     & Choosing the best tropical vacation location                    & P4                    \\\midrule
21             & \multirow{3}{*}{Health \& fitness}                  & Choosing an effective diet plan                                 & P1, P7                \\
22             &                                                     & Picking a reliable treadmill                                    & P3                    \\
23             &                                                     & Picking the best running shoes                                  & P8                    \\\midrule
24             & \multirow{3}{*}{Gifts \& special events}            & Choosing a birthday gift                                        & P5, P6                \\
25             &                                                     & Picking the right wedding venue                                 & P2                    \\
26             &                                                     & Picking an engagement ring                             & P2                    \\\midrule
27             & Parenting                                           & Choosing the best baby stroller                                 & P8                    \\\midrule
28             & Pets                                                & Choosing a breed of dog to adopt                                & P4  \\\bottomrule                 
\end{tabular}
}
\vspace{1mm}
\caption{Unfamiliar topics (organized by themes) that participants in the formative study reported encountering and exploring. Some topics were explored by multiple participants, such as ``Picking a smart home ecosystem'' and ``Choosing a reliable VPN provider.''}
\label{tab:selenite-formative-topics}
\end{table}

\clearpage
\pagebreak

\section{Study 1 Details}
\revision{
\subsection{Example web pages}\label{sec:apx-example-webpage}
Below in Figure \ref{fig:selenite-example-webpage}, we show a partial snapshot of a representative comparison article on the topic of ``best baby strollers.'' 
}

\begin{figure*}[h]
\vspace{-3mm}
\centering
\includegraphics[width=0.78\textwidth]{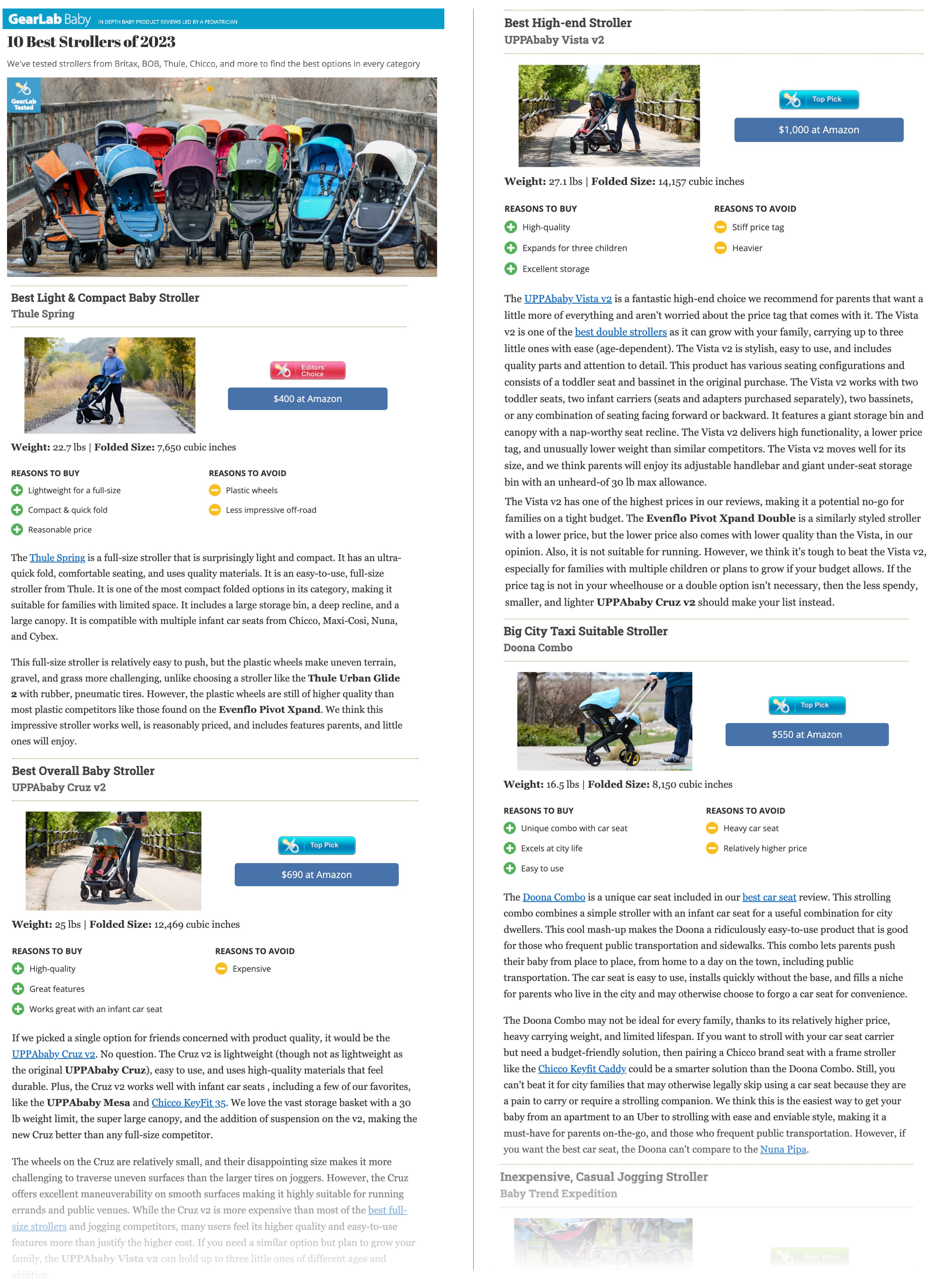}
\vspace{-5mm}
\caption{
\revision{Example comparison article on the topic of ``best baby strollers.'' Note that it only contains approximately $1/4$ of the article, which originally includes content about 10 baby strollers, along with other long-form commentaries. The article can be found online at \url{https://www.babygearlab.com/topics/getting-around/best-stroller}.}
}
\vspace{-3mm}
\label{fig:selenite-example-webpage}
\end{figure*}

\revision{
\subsection{Dataset of Groundtruth Criteria vs \systemname-retrieved Criteria}\label{sec:apx-groundtruth-criteria}
}

\newcommand{\groundTruthHit}{*\xspace}
\newcommand{\systemHit}{\^\xspace}

\def\arraystretch{1}
\begin{table*}[h]
\vspace{-5mm}
\centering
\resizebox{1\textwidth}{!}{%
\revision{
\begin{tabular}{
p{30mm}|
p{86mm}|
p{112mm}
}
\toprule
\textbf{Topic} & 
\textbf{Groundtruth Criteria} & \textbf{\systemname-retrieved Criteria} \\ 
\midrule

Best washing machines 
& 
Cleaning performance\groundTruthHit,
Fabric care\groundTruthHit,
Water usage\groundTruthHit,
Smart features\groundTruthHit,
Durability\groundTruthHit,
Water Temperature Options\groundTruthHit,
Noise level\groundTruthHit,
Load type\groundTruthHit,
Odor control\groundTruthHit,
Ease-of-use\groundTruthHit,
Price\groundTruthHit,
Aesthetics\groundTruthHit,
Power source\groundTruthHit,
Speed \& Cycle time\groundTruthHit,
Features\groundTruthHit,
Capacity\groundTruthHit,
Warranty \& Customer support\groundTruthHit,
Size\groundTruthHit,
Brand reputation\groundTruthHit 
(n = 19)
&
Capacity\systemHit,
Energy efficiency,
Water efficiency\systemHit,
Noise level\systemHit,
Durability\systemHit,
Cleaning performance\systemHit,
Cycle options\systemHit,
Spin speed\systemHit,
Price\systemHit,
Size\systemHit,
Brand reputation\systemHit,
User-friendliness / ease-of-use\systemHit,
Warranty\systemHit,
Features\systemHit,
Aesthetics\systemHit,
Cycle time\systemHit,
Load type\systemHit,
Maintenance,
Smart connectivity\systemHit,
Fabric care\systemHit,
Water Temperature Options\systemHit,
Power source\systemHit,
Child Safety Lock,
Odor Control\systemHit
~(n = 24)\\\midrule

Birthday gift ideas
&
Practicality\groundTruthHit,
Aesthetics\groundTruthHit,
Experience\groundTruthHit,
Price\groundTruthHit,
Age appropriateness\groundTruthHit,
Prerequisite,
Uniqueness\groundTruthHit,
Sentimental value\groundTruthHit,
Personalization\groundTruthHit,
Gender appropriateness\groundTruthHit,
Shipping\groundTruthHit 
(n = 11)
&
Personalization\systemHit,
Uniqueness\systemHit,
Practicality\systemHit,
Sentimental value\systemHit,
Presentation (wrapping),
Price\systemHit,
Creativity\systemHit,
Durability,
Functionality\systemHit,
Brand,
Age appropriateness\systemHit,
Gender appropriateness\systemHit,
Timeliness,
Accessibility (can get it in time)\systemHit,
Social norms,
Environment impact,
Size,
Experience\systemHit,
Aesthetic\systemHit,
Relevance,
Surprise factor 
(n = 21)
\\\midrule

Best hybrid app frameworks
&
Programming Language,
Code reusability\groundTruthHit,
Cross-platform compatibility\groundTruthHit,
Development time\groundTruthHit,
Ease of learning\groundTruthHit,
Third-Party Integration\groundTruthHit,
Popularity\groundTruthHit,
Features\groundTruthHit,
Testing and debugging\groundTruthHit,
Plugin availability\groundTruthHit,
Native features access\groundTruthHit,
Performance\groundTruthHit,
User interface\groundTruthHit,
Framework size\groundTruthHit,
Community support\groundTruthHit
(n = 15)
&
Cross-platform compatibility\systemHit,
Performance\systemHit,
User interface\systemHit,
Community support\systemHit,
Plugin availability\systemHit,
Development time\systemHit,
Maintenance,
Cost\systemHit,
Third-Party Integration\systemHit,
Security\systemHit,
Scalability\systemHit,
Platform Support\systemHit,
Ease of learning\systemHit,
Code reusability\systemHit,
Customization\systemHit,
Testing and debugging\systemHit,
App store compliance,
Innovation,
Popularity and Adoption\systemHit,
Framework size\systemHit,
Native features access\systemHit
~(n = 21)
\\\midrule

Best time tracking tools
&
Real-time tracking\groundTruthHit,
Editing ability\groundTruthHit,
Export and Invoice\groundTruthHit,
Accessibility\groundTruthHit,
Price\groundTruthHit,
OS platform\groundTruthHit,
Features\groundTruthHit,
Easy-of-use\groundTruthHit,
Tracking accuracy\groundTruthHit,
Simplicity\groundTruthHit,
Extensibility\groundTruthHit,
Automation\groundTruthHit,
Privacy\groundTruthHit,
Learning curve\groundTruthHit,
Intrusiveness,
Customization\groundTruthHit,
Interface\groundTruthHit,
Integration\groundTruthHit,
Focus work feature\groundTruthHit,
Project management\groundTruthHit
(n = 20)
&
User interface (ease-of-use, navigation)\systemHit,
Features\systemHit,
Compatibility\systemHit,
Reporting\systemHit,
Mobile access\systemHit,
Pricing\systemHit,
Customer support,
Customization\systemHit,
Security \& privacy\systemHit,
Time tracking accuracy\systemHit,
Time tracking methods (manual automatic timer)\systemHit,
Project management\systemHit,
Team management\systemHit,
Invoicing\systemHit,
Time off management\systemHit,
Analytics\systemHit,
User permissions,
Offline tracking,
Ease of setup\systemHit,
Work types\systemHit,
Team sizes
(n = 21)
\\\midrule

Deep learning frameworks
&
Programming language\groundTruthHit,
Data flow / graph structure\groundTruthHit,
Ease-of-use\groundTruthHit,
Support for product\groundTruthHit,
Support for research\groundTruthHit,
Visualization\groundTruthHit,
Release date,
OS platform\groundTruthHit,
Licensing\groundTruthHit,
Community support\groundTruthHit,
Growing speed (community),
Used-by,
Documentation\groundTruthHit,
Extensibility\groundTruthHit,
Developed-by,
Flexibility\groundTruthHit,
Scalability\groundTruthHit,
Speed / efficiency\groundTruthHit,
Performance\groundTruthHit,
Popularity\groundTruthHit,
Hardware support\groundTruthHit,
Supported network architecture\groundTruthHit,
Distributed compute\groundTruthHit,
Dependency\groundTruthHit,
CUDA usage\groundTruthHit
(n = 25)
&
Ease of use\systemHit,
Flexibility\systemHit,
Performance\systemHit,
Scalability\systemHit,
Community support\systemHit,
Compatibiilty (programming lamguage, hardware, OS, libraries)\systemHit,
Model accuracy (available pre-trained models)\systemHit,
Hardware support\systemHit,
Model interpretability (suppport for model understanding),
Model size,
Development speed,
Security,
Licensing\systemHit,
Data preprocessing\systemHit,
Model architecture\systemHit,
Documentation\systemHit,
Innovation (research)\systemHit,
Deployment\systemHit,
Debugging and profiling\systemHit,
Interoperability (integrate the model with other systems and applications)\systemHit
~(n = 20)
\\\midrule

Best sleeping bags
&
Temperature rating\groundTruthHit,
Comfort\groundTruthHit,
Weight\groundTruthHit,
Packed size\groundTruthHit,
Filling material\groundTruthHit,
Outside material\groundTruthHit,
Durability \groundTruthHit,
Price\groundTruthHit,
Easy to clean,
Space inside\groundTruthHit,
Shape\groundTruthHit,
Color\groundTruthHit,
Hooded or not\groundTruthHit,
Waterproofness\groundTruthHit,
Quality of zipper\groundTruthHit,
For extreme weather\groundTruthHit,
Easy setup,
Length\groundTruthHit,
Warranty\groundTruthHit
(n = 19)
&
Temperature rating\systemHit,
Weight\systemHit,
Insulation type\systemHit,
Comfort\systemHit,
Ventilation,
Size and fit\systemHit,
Packability\systemHit,
Durability\systemHit,
Water resistance\systemHit,
Zipper location,
Zipper quality\systemHit,
Price\systemHit,
Hood and collar design and availbility\systemHit,
Versatility (different environments and conditions)\systemHit,
Brand reputation,
Shape (mummy, rectangular, or semi-rectangular)\systemHit,
Length\systemHit,
Shell material\systemHit,
Lining material,
Warranty\systemHit,
Color options\systemHit
~(n = 21)
\\\midrule

Best air purifiers
&
Ease of use\groundTruthHit,
Coverage\groundTruthHit,
Noise\groundTruthHit,
Price\groundTruthHit,
Air quality indicator\groundTruthHit,
Speed settings\groundTruthHit,
Portability\groundTruthHit,
Design\groundTruthHit,
Smart features\groundTruthHit,
Warranty\groundTruthHit,
Customer reviews\groundTruthHit,
Customer support\groundTruthHit,
Filter type\groundTruthHit,
Sleep mode\groundTruthHit,
Remove smell\groundTruthHit,
Power consumption\groundTruthHit,
Performance\groundTruthHit,
Ozone safe\groundTruthHit,
Maintenance requirements\groundTruthHit,
Filter replacement indicator\groundTruthHit
(n = 20)
&
Filter type\systemHit,
Coverage area\systemHit,
CADR (Clean Air Delivery Rate),
Fan speed settings\systemHit,
Noise level\systemHit,
Energy efficiency\systemHit,
Additional features (such as ionizer, air quality sensors, auto mode, or remote control)\systemHit,
Design\systemHit,
Size,
Price\systemHit,
Brand reputation and customer reviews\systemHit,
effectiveness (dust, pollen, smoke, pet dander, and mold spores)\systemHit,
Maintenance and filter replacement\systemHit,
Odor elimination\systemHit,
Smart features\systemHit,
Filter indicator\systemHit,
Warranty\systemHit,
User-friendly controls\systemHit,
Portability\systemHit,
Ozone emissions (high levels can be harmful to health)\systemHit,
Customer support\systemHit,
Sleep mode\systemHit,
Child lock,
Certifications
(n = 24)
\\\midrule

Best robot vacuums
&
Performance\groundTruthHit,
Pet hair handling\groundTruthHit,
Battery life\groundTruthHit,
Reliability\groundTruthHit,
Ease of use\groundTruthHit,
Price\groundTruthHit,
Design\groundTruthHit,
Functionality\groundTruthHit,
Dust bin size\groundTruthHit,
Customer support\groundTruthHit,
Suction power\groundTruthHit,
Ability to handle different floor types\groundTruthHit,
Noise level\groundTruthHit,
Compactness\groundTruthHit,
Navigation capabilities\groundTruthHit,
Maintenance requirements\groundTruthHit,
Auto-recharge\groundTruthHit,
Smart home compatibility\groundTruthHit,
Scheduling capabilities\groundTruthHit,
Boundary control features\groundTruthHit,
Voice control\groundTruthHit,
Effectiveness in corners and edges\groundTruthHit,
Automatic dirt disposal\groundTruthHit
(n = 23)
&
Cleaning performance\systemHit,
Navigation and mapping\systemHit,
Battery life and charging\systemHit,
Noise level\systemHit,
Dustbin capacity\systemHit,
Smart features and connectivity\systemHit,
Edge cleaning and corner reach\systemHit,
Maintenance and filter replacement\systemHit,
Price and value for money\systemHit,
Customer reviews and ratings,
Runtime and automatic docking\systemHit,
Scheduling and automation\systemHit,
Filtration system (the efficiency of the vacuum's filtration system in capturing small particles and allergens),
Suction power\systemHit,
Size and design\systemHit,
Multi-floor cleaning\systemHit,
Virtual walls and boundary setting\systemHit,
Pet hair performance\systemHit,
Durability and reliability\systemHit,
Brand reputation and customer support\systemHit,
Accessories and additional features\systemHit,
User-friendly interface \& control\systemHit,
Warranty and after-sales service,
Mopping capabilities,
Remote control options (e.g., voice)\systemHit,
Energy efficiency,
Auto-emptying capabilities\systemHit,
Integration with smart home systems\systemHit
~(n = 28)
\\\midrule

Best baby strollers
&
Safety features\groundTruthHit,
Comfort for baby\groundTruthHit,
Durability\groundTruthHit,
Foldability\groundTruthHit,
Steer and maneuver\groundTruthHit,
Storage space\groundTruthHit,
Color options\groundTruthHit,
For twins\groundTruthHit,
Compatibility with car seats\groundTruthHit,
Customer support\groundTruthHit,
Reviews\groundTruthHit,
Ease of cleaning\groundTruthHit,
Expandability for growing child\groundTruthHit,
Suspension system\groundTruthHit,
Terrain adaptability\groundTruthHit,
Included accessories (like cup holders, trays, etc.)\groundTruthHit,
Canopy and weather protection\groundTruthHit,
Height and angle of handlebars\groundTruthHit,
Reversible seat\groundTruthHit,
Price\groundTruthHit,
Weight and size\groundTruthHit,
Lockable swivel wheels\groundTruthHit
(n = 22)
&
Safety (secure harness, sturdy construction, and reliable brakes)\systemHit,
Comfort\systemHit,
Maneuverability\systemHit,
Durability\systemHit,
Storage\systemHit,
Folding mechanism and ability\systemHit,
Weight and size\systemHit,
Versatility (terrain)\systemHit,
Price\systemHit,
Ease of cleaning \& maintenance\systemHit,
Design and Style\systemHit,
Suspension\systemHit,
Canopy \& UV and weather protection\systemHit,
Reversible seat\systemHit,
Adjustable handlebar (to accommodate parents of different heights and provide comfortable pushing)\systemHit,
Brake \& locking system (ensure the stroller stays in place when needed)\systemHit,
Travel system compatibility\systemHit,
Accessories\systemHit,
Customer reviews and ratings\systemHit,
Adjustablility (to accommodate baby growth)\systemHit,
Adjustable seat height (to accommodate different table heights),
Warranty \& customer support\systemHit,
Easy assembly,
Accommodation for twins\systemHit
~(n = 24)
\\\midrule

Best tropical vacation spots
&
Weather\groundTruthHit,
Cost\groundTruthHit,
Beach quality\groundTruthHit,
Cultrual experiences\groundTruthHit,
Outdoor activities\groundTruthHit,
Ease of travel\groundTruthHit,
Hotel\groundTruthHit,
Scenery\groundTruthHit,
Child-friendly\groundTruthHit,
Shopping opportunities\groundTruthHit,
Undertanding of language\groundTruthHit,
Food\groundTruthHit,
LGBTQ+ friendly,
Bars and clubs\groundTruthHit,
Safety\groundTruthHit
(n = 15)
&
Weather\systemHit,
Beaches\systemHit,
Activities \& Adventure opportunities\systemHit,
Natural beauty\systemHit,
Accommodation\systemHit,
Safety\systemHit,
Transportation \& Accessibility (Ease of travel and transportation options)\systemHit,
Culture \& history\systemHit,
Dining options\systemHit,
Cost \& Value for money\systemHit,
Nightlife (bars, clubs, or live music venues)\systemHit,
Reviews and recommendations,
Family-friendly\systemHit,
Wildlife,
Local hospitality,
Shopping\systemHit,
Sustainability (Commitment to eco-tourism practices, conservation efforts, and protection of the environment),
Communication \& Language barrier\systemHit,
Accessibility for people with disabilities
(n = 19)
\\\bottomrule
\end{tabular}%
}
}
\caption{\revision{Dataset of Groundtruth Criteria vs \systemname-retrieved Criteria. Groundtruth criteria that are retrieved by \systemname are marked with a ``\groundTruthHit'', while \systemname-retrieved criteria that appear in the Groundtruth set are marked with a ``\systemHit''.}}
\label{tab:selenite-groundtruth-criteria}
\vspace{-9mm}
\end{table*}

\pagebreak


\section{Topics Explored in Study 3}

\begin{table*}[h]
\centering
\resizebox{1\textwidth}{!}{
\begin{tabular}{c|p{72mm} || c|p{72mm}}
\toprule
\textbf{Participant} & \textbf{Topics}                                            & \textbf{Participant} & \textbf{Topics}                              \\\midrule\midrule
\multirow{3}{*}{P1}  & Choosing a high-quality digital camera                     & \multirow{3}{*}{P5}  & Picking the best robot vacuum                \\
                     & Choosing the best air purifier                             &                      & Picking a drone                              \\
                     & Picking a suitable hand truck                              &                      & Choosing the best e-reader                   \\\midrule
\multirow{3}{*}{P2}  & Selecting the best washing machine                         & \multirow{3}{*}{P6}  & Picking the right refrigerator               \\
                     & Choosing a reliable VPN provider &                      & Choosing the best barbecue grill             \\
                     & Deciding on unique thank-you gifts                         &                      & Choosing the best tropical vacation location \\\midrule
\multirow{3}{*}{P3}  & Selecting the best mattress                                & \multirow{3}{*}{P7}  & Choosing an effective diet plan              \\
                     & Choosing the best city bike                                &                      & Choosing a breed of dog to adopt             \\
                     & Choosing a birthday gift                                   &                      & Selecting a suitable SUV                     \\\midrule
\multirow{3}{*}{P4}  & Selecting a secure password manager                        & \multirow{3}{*}{P8}  & Picking a reliable treadmill                 \\
                     & Choosing the best tropical vacation location               &                      & Picking the right wedding venue              \\
                     & Choosing a hybrid app framework                            &                      & Choosing the best skiing venue    \\\bottomrule          
\end{tabular}
}
\caption{Topics that participants in study 3 explored.}
\label{tab:selenite-study-3-topics}
\end{table*}
\definecolor{stepcolor}{rgb}{0.96, 0.73, 1.0}
\newcommand{\stepIndicator}[1]{
\texttt{
\colorbox{stepcolor}{Step #1}\xspace
}}

\definecolor{cgenerate}{HTML}{3B4D73}
\definecolor{cprompt}{HTML}{3c4043}

\newcommand{\tprompt}[1]{
\noindent{\color{cprompt}\texttt{#1}}\xspace
}
\newcommand{\tgenerate}[1]{\noindent{\color{cgenerate}\texttt{#1}}\xspace
}

\newcommand{\systemnamemessage}{\noindent\texttt{[System Message]}\xspace}
\newcommand{\usermessage}{\noindent\texttt{[User Message]}\xspace}
\newcommand{\assistantmessage}{\noindent\texttt{[Assistant Message]}\xspace}

\newcommand{\stepBreakline}{
\vspace{-2mm}
\noindent\rule{\textwidth}{0.2pt}
}

\section{\gptFour Prompts used in \systemname}\label{sec:gpt-prompts}

Here, we outline the techniques we employed to guide the \gptFour model, developed by OpenAI \cite{openai_gpt-4_2023}, within the context of \systemname. If not explicitly stated, the temperature of the model is set to 0.3 for a balance between consistency and creativity.

If not explicitly stated, the initial \texttt{[System Message]}\footnote{The system message helps set the behavior of the model response, i.e., the \texttt{[Assistant Message]}. However, as stated by OpenAI: ``... note that the system message is optional and the model's behavior without a system message is likely to be similar to using a generic message such as `You are a helpful assistant' ...'' (\url{https://platform.openai.com/docs/guides/gpt/chat-completions-api})} was set as the following:


\systemnamemessage

\tprompt{You are a helpful assistant that performs content analysis according to user requests. Follow the user's requirements carefully and to the letter.}


\subsection{Obtaining topic from a web page}\label{sec:gpt-prompt-get-topic}

The prompt that we used to obtain a concise topic given a web page (with \texttt{[title]} and \texttt{[content of the first few paragraphs]}) is a two-step prompt:

\noindent\rule{\textwidth}{1pt}
\usermessage \stepIndicator{1}

\tprompt{Given the following information of an article:}

\tprompt{Title:}
\tprompt{[title]}

\tprompt{First few paragraphs:}
\tprompt{[content of the first few paragraphs]}

\tprompt{What is this article about?}
\vspace{5mm}

\assistantmessage

\tgenerate{This article is about ...}
\vspace{5mm}

\usermessage \stepIndicator{2}

\tprompt{I want to find articles similar to this one in terms of the general topic. What should I search for? Output one search phrase (in double quotes).}

\noindent\rule{\textwidth}{1pt}

Additionally, we set the \texttt{n} parameter to 10, thereby instructing \gptFour to produce 10 simultaneous responses. Subsequently, we determined the most commonly occurring one among these 10 as the \textbf{topic} for the article.  While it could be assumed that a higher value for \texttt{n} would result in a lengthier response time from the model, our observations indicate that such delay is practically insignificant.

\subsection{Obtaining options from a web page}\label{sec:gpt-prompt-get-options}

The prompt that we used to obtain extract options from a given web page (with \texttt{[title]} and \texttt{[content of the web page]}) is a two-step prompt:

\noindent\rule{\textwidth}{1pt}



\usermessage \stepIndicator{1}

\tprompt{Given the following information of an article:}

\tprompt{~}

\tprompt{Title:}
\tprompt{[title]}

\tprompt{~}

\tprompt{Is the article likely to be discussing one or more aspects of "one specific option" (e.g., a single javascript framework, for example, React, or a single baby stroller option, or a specific Airbnb listing) or "multiple options/topics"? Output in the following format:}

\tprompt{Reasoning: your reasoning process.}

\tprompt{Verdict: "one specific option / multiple options"}
\vspace{5mm}

\assistantmessage

\tgenerate{Reasoning: model's reasoning process...}

\tgenerate{Verdict: "one specific option / multiple options"}
\vspace{5mm}

\usermessage \stepIndicator{2}

\tprompt{Now, given the content of the article below, what is/are the options?}

\tprompt{Content:}

\tprompt{[content of the web page]}

\tprompt{~}

\tprompt{Output should be in the following format: ["option\_1", "option\_2", ...]}

\noindent\rule{\textwidth}{1pt}

\subsection{Obtaining commonly-considered criteria from a web page}\label{sec:gpt-prompt-get-criteria}

The prompt that we used to obtain a set of commonly considered criteria resembles something like the following (given a \texttt{[topic]}):

\noindent\rule{\textwidth}{1pt}
\usermessage \stepIndicator{1: Ask for an initial set of criteria}

\tprompt{What are some common aspects, criteria, or dimensions that people consider on the topic of [topic]? Note that the criteria should be **most relevant to the topic**, **frequently considered**, and can **cover a broad range of perspectives**. Output should be a single bulleted list in the format of:}

\tprompt{~}

\tprompt{- Criterion: short description.}

\tprompt{~}

\tprompt{Do not output anything else.}
\vspace{5mm}

\assistantmessage

\tgenerate{- [Criterion 1]: [Short description]}

\tgenerate{- [Criterion 2]: [Short description]}

\tgenerate{- [Criterion 3]: [Short description]}

\tgenerate{...}
\vspace{5mm}

\usermessage \stepIndicator{2+: Ask for additional criteria until we get around 20.}

\tprompt{Give me five more that are different from, more diverse than, and possibly as important as the ones listed above. Output in the same format.}

\noindent\rule{\textwidth}{1pt}

\subsection{Obtaining detailed analysis of text content}\label{sec:gpt-prompt-get-detailed-analysis} 

The prompts that we used to obtain a detailed analysis of text content given the \texttt{[text content]}, the list of \texttt{NLI criteria}, and the list of \texttt{[options]} on the corresponding web page is two-fold:

First, we ask \gptFour to extract phrases from the content that describes a given criterion as well as determine each extracted phrase's sentiment with respect to the criterion:

\noindent\rule{\textwidth}{1pt}

\usermessage 

\tprompt{Given the following **content** and list of **criteria**:}

~

\tprompt{**Content**:}

~

\tprompt{[content]}

~

\tprompt{**Criteria (with definitions)**:}

\tprompt{- [NLI Criterion 1]: [description]}

\tprompt{- [NLI Criterion 2]: [description]}

\tprompt{...}

~ 

\tprompt{For each criterion: 1) extract **every possible** utterance that **mentions** or **explicitly describes** that criterion from the content 2) perform sentiment analysis to determine if the utterance is "positive", "neutral", or "negative" with respect to that criterion. Remember to use the **exact same words** from the content. Do not paraphrase!}

~ 

\tprompt{Output must follow the format below:}

~

\tprompt{\#\# criterion\_1\_name}

\tprompt{- "extracted\_sentence\_or\_phrase\_1" -> positive,}

\tprompt{- "extracted\_sentence\_or\_phrase\_2" -> neutral,}

\tprompt{\#\# criterion\_2\_name}

\tprompt{NONE FOUND}

\tprompt{\#\# criterion\_3\_name}

\tprompt{- "extracted\_sentence\_or\_phrase\_1" -> neutral,}

\tprompt{- "extracted\_sentence\_or\_phrase\_2" -> negative,}

\tprompt{- "extracted\_sentence\_or\_phrase\_3" -> positive,}

\noindent\rule{\textwidth}{1pt}

Second, we ask \gptFour to label each extracted phrase with a possible \texttt{[option]} on the web page (we framed options as ``subjects'' of a phrase to achieve a better empirical performance):

\noindent\rule{\textwidth}{1pt}

\usermessage 

\tprompt{Given the following **content** and the **phrases** extracted from the content below:}

~

\tprompt{**Content**:}

~

\tprompt{[content]}

~

\tprompt{**Extracted phrases**:}

\tprompt{- "extracted\_phrase\_1"}

\tprompt{- "extracted\_phrase\_2"}

\tprompt{- "extracted\_phrase\_3"}

\tprompt{...}

~

\tprompt{For each phrase, determine the **subject** of the phrase based on the **content**. Possible subjects are: [option\_1, option\_2, option\_3, ...] Say "N/A" if you cannot determine the subject. Output should be in the following format:}

~

\tprompt{"extracted phrase 1" -> "subject" or "N/A"}

\tprompt{"extracted phrase 2" -> "subject" or "N/A"}

\tprompt{...}

\noindent\rule{\textwidth}{1pt}

\revision{

\section{Other Implementation Details}\label{sec:addition-implementation-details}

We leverage \gptFour for several use cases in \systemname, and encountered several challenges\footnote{We have documented the specific prompt designs for those tasks in section \ref{sec:gpt-prompts}, and only discuss a series of challenges we experienced while interacting with the \gptFour API here in this section.}:  
First, due to the limited context window size of \gptFour (8192 tokens or approximately 6100 English words), we occasionally need to divide the entire text content of a web page into smaller chunks and run parallel queries to extract options.
As of July 2023, we don't have access to the version of \gptFour with a 32k context window, which would significantly reduce the need for chunking and parallel queries.
Second, unfortunately, there are occasions when the \gptFour model becomes overloaded with requests or takes an exceptionally long time to respond. To mitigate these problems and provide uninterrupted user experience to \systemname users, we have employed the following two approaches: 1) \emph{Dual API requests}: We send two identical requests using separate API keys simultaneously. We prioritize the response that returns first with valid information, indicating that it is not an error and contains the requested information from the prompt; 2) \emph{Graceful error handling \& retry}: In the event of an error, we introduce a random delay (ranging from 1 to 5 seconds) before retrying the request. We repeat this retry process for up to 5 attempts, allowing sufficient opportunity for a successful response. 
Note that these issues are attributable, in part, to the current limited beta status of \gptFour. Consequently, it is uncertain whether these issues will persist in the future. Nevertheless, we delve into them here to provide a comprehensive and accurate accounting of our experience interacting with the API.

To efficiently perform natural language inference (NLI) during the analysis of article content to produce per-paragraph summaries and annotations of options and criteria, we experimented with both the \texttt{roberta-large-mnli} and \texttt{bart-large-mnli} models that are fine-tuned for multi-genre natural language inference (MNLI) tasks\footnote{These two models are considered to be able to achieve state-of-the-art performance as of June 2023.}, and ended up using the latter for its better performance in our informal testing. In addition, we implemented a REST API service that the Chrome extension can query on demand. To decrease model inference time and ensure a smooth user experience, we ran the service on multiple Google Cloud virtual machines with NVIDIA L4 GPUs. 

}

\end{document}